\documentstyle[12pt]{article}
\newcommand{\bb}{\bibitem}
 
\begin{document}
 
\begin{titlepage}
\begin{flushright}
       {\bf UK/97-23, ADP-97-55/T281}  \\
Dec. 1997   \\
       hep-ph/9712483 \\
\end{flushright}
\begin{center}
 
{\bf {\LARGE Lattice Calculation of the Strangeness Magnetic Moment of the
Nucleon}}
 
\vspace{1cm}
 
{\bf S.J.\ Dong$^a$, K.F.\ Liu$^a$, and
A.G.\ Williams$^b$} \\
[0.5em]
 {\it  $^a$ Dept. of Physics and Astronomy, Univ. of Kentucky,
            Lexington, KY 40506\\
       $^b$ Special Research Cneter for the Subatomic Structure of Matter
            and Department of Physics and Mathematical Physics,
            University of Adelaide, Australia 5005 }
 
\end{center}
 
\vspace{0.4cm}
 
\begin{abstract}
 
We report on a lattice QCD calculation
of the strangeness magnetic moment of the nucleon. Our result is
$G_M^s(0) =  - 0.36 \pm 0.20 $. The sea contributions from the u and d
quarks are about 80\% larger. However, they cancel to a large extent
due to their electric charges, resulting in a smaller net sea contribution
of $ - 0.097 \pm 0.037 \mu_N$ to the nucleon magnetic moment. As far as the
neutron to proton magnetic moment ratio is concerned, this sea contribution
tends to cancel out the cloud-quark effect from the Z-graphs and result
in a ratio of $ -0.68 \pm 0.04$ which is close to the SU(6) relation and
the experiment.
The strangeness Sachs electric mean-square radius $\langle r_s^2\rangle_E$
is found to be small and negative and the total sea contributes
substantially to the neutron electric form factor.

\bigskip
 
 
\end{abstract}
 
\vfill
 
\end{titlepage}

The strangeness content of the nucleon has been a topic of considerable
recent interest for a variety of reasons. The studies of nucleon
spin structure functions in polarized deep inelastic
scattering experiments at CERN and SLAC \cite{DIS}, combined with
neutron and hyperon $\beta$ decays, have turned up a surprisingly
large and negative polarization from the strange quark. In addition,
there is a well-known long-standing discrepancy between the
pion-nucleon sigma term
extracted from the low energy pion-nucleon scattering~\cite{gls91}
and that from the octect baryon masses~\cite{cheng76}. This
discrepancy can be reconciled if a significant $\bar{s}s$ content
in the nucleon~\cite{gl82,cheng76} is admitted.
 
This naturally leads to the question that if the strange quarks can
contribute substantially in the axial-vector and scalar current matrix
elements, how important are they in other matrix elements involving the
vector, pseudoscalar, and tensor currents.
The case of the vector current matrix element
$\langle N |\bar{s}\gamma_{\mu} s|N\rangle$
is especially interesting. If the strange magnetic moment (m.\,m.) is
large, it is likely to spoil the nice SU(6) prediction of the
neutron to proton m.\,m. ratio of $- 3/5$ which lends credence
to the valence quark picture. On the other hand, if it is small one would
like to understand
why it should be different from the axial-vector and scalar cases.

To address some of these issues, an experiment to
measure the neutral weak magnetic form factor $G^Z_M$
via elastic parity-violating electron scattering at backward angles
was recently carried out
by the SAMPLE collaboration~\cite{SAMPLE97}. The strangeness magnetic form
factor is then obtained by subtracting out the nucleon magnetic form
factors $G^p_M$ and $G^n_M$. The reported value is
$G^s_M(Q^2=0.1$GeV$^2)= +0.23\pm 0.37\pm 0.15\pm 0.19 $, where the last
error is due to an uncertainty associated with axial radiative corrections.
Future experiments have the promise of tightening the errors and isolate
the radiative corrections so that we can hope to have a well-determined
value and sign for $G^s_M(0)$.
 
Theoretical predictions of $G^s_M(0)$ vary widely. The values
from various models and analyses vary from $ -0.75\pm 0.30 $
in a QCD equalities analysis~\cite{lei96} to $+ 0.37 $ in an $SU(3)$
chiral bag model\cite{hpm97}. While a few give positive values
~\cite{hpm97,gi97}, most model predictions are negative with a
typical range of $-0.25$ to $-0.45$~\cite{lei96,jaf89,psw91,
mb94,fnj94,hmd96,cbk96}. Summaries of these predictions can be found in
Refs.\ \cite{lei96,cbk96}. A similar situation exists for the
strangeness electric mean-square radius $\langle r_s^2\rangle_E$.
A number of the predictions are positive~\cite{jaf89,hmd96}, while the
others are negative~\cite{gi97,psw91,mb94,fnj94,cbk96}.
Elastic $\vec{e}\; p$ and $\vec{e}\;{}^{4}He$ parity-violation experiments
are currently planned at TJNAF~\cite{pve91} to
measure the asymmetry $A_{LR} $ at forward angles to extract
the strangeness electric mean-square radius. Hopefully, they will settle
the issue of the sign of $\langle r_s^2\rangle_E$.
 
In view of the large spread of theoretical predictions for both
$G^s_M(0)$ and $\langle r_s^2\rangle_E$ and in view of the fact
that the experimental errors on $G^s_M(0)$ are still large, it is clearly
important to perform a lattice calculation of this quantity
since this is a first principles theoretical approach. We present
our calculation from lattice QCD in the hope that this will shed some light
on these quantities. Our previous results on flavor-singlet quantities
which involve the so-called ``disconnected insertions'' (DI)
for the sea quarks in addition to the ``connected insertions''
(CI) for the valence and cloud quarks~\cite{dll95,dll96} reveal that
the sea quark contribution to the flavor-singlet $g_A^0$ from the
DI is negative and the magnitude large enough (e.g. the
strangeness polarization $\Delta s = 0.12 \pm 0.01$) to cancel
the positive CI contribution to a large extent. This results in
a small $g_A^0$ at $0.25 \pm 0.12$, which is
in agreement with the experimental results~\cite{DIS}. Similarly, the
calculated ratio $y = \langle N|\bar{s}s|N\rangle/\langle N|\bar{u}u +
\bar{d}d |N\rangle = 0.36 \pm 0.03$~\cite{dll96} gives the right amount of
strangeness content to resolve the $\pi N \sigma$ puzzle we alluded to
earlier. Given these reasonably successful estimates of strangeness
in the axial-vector and scalar channels, we feel that it should yield
meaningful results in the vector current as well. In particular, we would
like to understand why the SU(6) valence quark picture fails badly in the
flavor-singlet axial-vector and scalar cases and yet gives an apparently good
prediction in the neutron to proton m.\,m. ratio -- a yet unresolved
puzzle in low-energy hadron physics.
 
The lattice formulation of the electromagnetic form factors has been given
in detail in the past~\cite{dwl90,wdl92}. Here, we shall concentrate on the
DI contribution, where the strangeness current contributes.
In the Euclidean formulation, the Sachs electromagnetic form factors
can be obtained by the combination of two- and three-point functions
\begin{equation} \label{twopt}
G_{NN}^{\alpha\alpha}(t,\vec{p}) = \sum_{\vec{x}}e^{-i\vec{p}\cdot
\vec{x}}  \langle 0| \chi^\alpha(x) \bar{\chi}^\alpha(0) |0 \rangle
\end{equation}
\begin{equation} \label{threept}
G_{NV_{\mu}N}^{\alpha\beta}(t_f,\vec{p},t,\vec{q})=
\sum_{\vec{x}_f,\vec{x}} e^{-i\vec{p}\cdot\vec{x}_f
  +i\vec{q}\cdot\vec{x}} \langle 0| \chi^\alpha(x_f) V_\mu(x)
  \bar{\chi}^\beta(0) |0 \rangle ,
\end{equation}
where $\chi^\alpha$ is the nucleon interpolating field and $V_{\mu}(x)$
the vector current.
With large Euclidean time separation, i.e. $t_f - t >> a$ and $t >> a$,
where $a$ is the lattice spacing,
\begin{eqnarray}
\frac{\Gamma_i^{\beta\alpha}G_{NV_jN}^{\alpha\beta}(t_f,\vec{0},t,
\vec{q})} {G_{NN}^{\alpha\alpha}(t_f,\vec{0})}
\frac{G_{NN}^{\alpha\alpha}(t,\vec{0})}{G_{NN}^{\alpha\alpha}
(t,\vec{q})} \longrightarrow  \frac{\varepsilon_{ijk}q_k}{E_q + m}
 G_M(q^2),  \label{mff} \\
\frac{\Gamma_E^{\beta\alpha}G_{NV_4N}^{\alpha\beta}(t_f,\vec{0},t,
\vec{q})} {G_{NN}^{\alpha\alpha}(t_f,\vec{0})}
\frac{G_{NN}^{\alpha\alpha}(t,\vec{0})}{G_{NN}^{\alpha\alpha}
(t,\vec{q})} \longrightarrow   G_E(q^2), \label{eff}
\end{eqnarray}
where $\Gamma_i=\left(\begin{array}{cc}
                 \sigma_i  & 0 \\
                 0 & 0\end{array}\right)$,
and $\Gamma_E=\left(\begin{array}{cc}
                 1  & 0 \\
                 0 & 0\end{array}\right)$.
 
We shall use the conserved current from the Wilson action which, being
point-split, yields slight variations
on the above forms and these are given in Ref.~\cite{dwl90,wdl92}.
Our 50 quenched gauge configurations were generated on a $16^3 \times 24$
lattice at $\beta = 6.0$.
In the time direction,
Dirichlet boundary conditions were imposed on the quarks to provide
larger time separations than available with periodic boundary
conditions. We also averaged over the directions of equivalent
lattice momenta in each configuration; this has the desirable effect of
reducing error bars.
Numerical details of this procedure are given in Refs. \cite{wdl92,ldd95}.
The dimensionless nucleon masses $M_N a$ for
$\kappa = 0.154$, 0.152, and 0.148 are 0.738(16), 0.882(12), and
1.15(1) respectively. The corresponding dimensionless
pion masses $m_{\pi} a$ are
0.376(6), 0.486(5), and 0.679(4). Extrapolating the nucleon and
pion masses to the chiral limit we determine $\kappa_c = 0.1567(1)$
and the dimensionless nucleon mass at the chiral limit to be 0.547(14).
Using the nucleon mass to set the scale, which we believe to be
appropriate for studying nucleon properties~\cite{ldd94,ldd95,dll95,dll96},
the lattice spacing $a^{-1} = 1.72(4)$ GeV is determined. The
three $\kappa's$ then correspond to quark masses of about 120, 200,
and 360 MeV respectively.
 
The strangeness current $\bar{s}\gamma_{\mu}s$ contribution appears in the
DI only. In this case, we sum up the current insertion
time slice $t$ from the nucleon source to the sink in Eqs.(\ref{mff})
and (\ref{eff}) to gain statistics~\cite{dll95,dll96}. This
leads to ${\rm const} + t_f G_{E,{\rm dis}} (q^2)$ for Eq. (\ref{eff}). For
Eq. (\ref{mff}), we average over the three spatial components
$\bar{s}\gamma_i s$ and obtain ${\rm const} + t_f \frac{|\vec{q}|}{E_q + m}
G_{M,{\rm dis}} (q^2)$. Similar to our previous studies of $\Delta s$
~\cite{dll95} and $\langle N|\bar{s}s|N\rangle$~\cite{dll96}, we use
300 complex $Z_2$ noises~\cite{dl94} and 100 gauge configurations to
calculate the sea quark contribution (DI) with $\kappa = 0.148, 0.152$
and 0.154. In calculating the strange current,
we have considered the correlation between the quark loop with
$\kappa_s = 0.154$
and the valence quarks at $\kappa_v = 0.148, 0.152$, and 0.154.
The ratio in Eq. (\ref{mff}) with the sum in $t$ and average in $V_i$, which
leads to the expression  ${\rm const} + t_f G_{M,{\rm dis}} (q^2)$,
is plotted in Fig. 1 as a function of $t_f$ for $|\vec{q}| = 2\pi/La$. Then
$G_{M,{\rm dis}} (q^2)$ from the DI is obtained from fitting the slopes
in the region $t_f \ge 8$ where the nucleon is isolated from its
excited states with the correlation among the time slices taken
into account~\cite{dll95}. The resultant straight-line fits covering
the ranges of $t_f$ with the minimum $\chi^2$ are plotted in Fig. 1.
Finally, the errors on the fit, also shown in the figure, are
obtained by jackknifing the procedure.
To obtain the physical $G_M^s(q^2)$, we extrapolate the valence
quarks to the chiral limit while keeping the sea quark in the loop
at the strange quark
mass (i.e. $\kappa_s$ = 0.154). It has been shown in the chiral
perturbation theory with a kaon loop that $G_M^s(0)$ is proportional to
$m_K$, the kaon mass~\cite{gss88}. Thus, we extrapolate
with the form $C + D \sqrt{\hat{m} + m_s}$ where $\hat{m}$ is the
average u and d quark mass and $m_s$ the strange quark mass to
reflect the $m_K$ dependence. This is the same form adopted for
extracting $\langle N|\bar{s}s|N\rangle$ in Ref.~\cite{dll96},
which also involves a kaon loop in the chiral perturbation theory.
 
\begin{figure}[h]
\[
\hspace*{-0.8in}
\setlength{\unitlength}{0.240900pt}
\ifx\plotpoint\undefined\newsavebox{\plotpoint}\fi
\sbox{\plotpoint}{\rule[-0.200pt]{0.400pt}{0.400pt}}%
\begin{picture}(825,629)(0,0)
\font\gnuplot=cmr10 at 10pt
\gnuplot
\sbox{\plotpoint}{\rule[-0.200pt]{0.400pt}{0.400pt}}%
\put(176.0,168.0){\rule[-0.200pt]{4.818pt}{0.400pt}}
\put(154,168){\makebox(0,0)[r]{-1}}
\put(741.0,168.0){\rule[-0.200pt]{4.818pt}{0.400pt}}
\put(176.0,277.0){\rule[-0.200pt]{4.818pt}{0.400pt}}
\put(154,277){\makebox(0,0)[r]{-0.5}}
\put(741.0,277.0){\rule[-0.200pt]{4.818pt}{0.400pt}}
\put(176.0,386.0){\rule[-0.200pt]{4.818pt}{0.400pt}}
\put(154,386){\makebox(0,0)[r]{0}}
\put(741.0,386.0){\rule[-0.200pt]{4.818pt}{0.400pt}}
\put(176.0,495.0){\rule[-0.200pt]{4.818pt}{0.400pt}}
\put(154,495){\makebox(0,0)[r]{0.5}}
\put(741.0,495.0){\rule[-0.200pt]{4.818pt}{0.400pt}}
\put(176.0,113.0){\rule[-0.200pt]{0.400pt}{4.818pt}}
\put(176,68){\makebox(0,0){0}}
\put(176.0,541.0){\rule[-0.200pt]{0.400pt}{4.818pt}}
\put(348.0,113.0){\rule[-0.200pt]{0.400pt}{4.818pt}}
\put(348,68){\makebox(0,0){5}}
\put(348.0,541.0){\rule[-0.200pt]{0.400pt}{4.818pt}}
\put(520.0,113.0){\rule[-0.200pt]{0.400pt}{4.818pt}}
\put(520,68){\makebox(0,0){10}}
\put(520.0,541.0){\rule[-0.200pt]{0.400pt}{4.818pt}}
\put(692.0,113.0){\rule[-0.200pt]{0.400pt}{4.818pt}}
\put(692,68){\makebox(0,0){15}}
\put(692.0,541.0){\rule[-0.200pt]{0.400pt}{4.818pt}}
\put(176.0,113.0){\rule[-0.200pt]{140.926pt}{0.400pt}}
\put(761.0,113.0){\rule[-0.200pt]{0.400pt}{107.923pt}}
\put(176.0,561.0){\rule[-0.200pt]{140.926pt}{0.400pt}}
\put(468,23){\makebox(0,0){$t_f$}}
\put(468,606){\makebox(0,0){$\kappa_v=0.154$}}
\put(245,277){\makebox(0,0)[l]{{\tiny $M=-0.12(5)$}}}
\put(245,233){\makebox(0,0)[l]{{\tiny $\chi^2$=0.02}}}
\put(176.0,113.0){\rule[-0.200pt]{0.400pt}{107.923pt}}
\put(210,386){\circle*{18}}
\put(245,382){\circle*{18}}
\put(279,381){\circle*{18}}
\put(314,389){\circle*{18}}
\put(348,399){\circle*{18}}
\put(382,394){\circle*{18}}
\put(417,398){\circle*{18}}
\put(451,402){\circle*{18}}
\put(486,415){\circle*{18}}
\put(520,404){\circle*{18}}
\put(555,377){\circle*{18}}
\put(589,339){\circle*{18}}
\put(623,331){\circle*{18}}
\put(658,309){\circle*{18}}
\put(692,281){\circle*{18}}
\put(727,261){\circle*{18}}
\put(210,386){\usebox{\plotpoint}}
\put(200.0,386.0){\rule[-0.200pt]{4.818pt}{0.400pt}}
\put(200.0,386.0){\rule[-0.200pt]{4.818pt}{0.400pt}}
\put(245.0,378.0){\rule[-0.200pt]{0.400pt}{1.927pt}}
\put(235.0,378.0){\rule[-0.200pt]{4.818pt}{0.400pt}}
\put(235.0,386.0){\rule[-0.200pt]{4.818pt}{0.400pt}}
\put(279.0,373.0){\rule[-0.200pt]{0.400pt}{3.613pt}}
\put(269.0,373.0){\rule[-0.200pt]{4.818pt}{0.400pt}}
\put(269.0,388.0){\rule[-0.200pt]{4.818pt}{0.400pt}}
\put(314.0,372.0){\rule[-0.200pt]{0.400pt}{8.431pt}}
\put(304.0,372.0){\rule[-0.200pt]{4.818pt}{0.400pt}}
\put(304.0,407.0){\rule[-0.200pt]{4.818pt}{0.400pt}}
\put(348.0,369.0){\rule[-0.200pt]{0.400pt}{14.454pt}}
\put(338.0,369.0){\rule[-0.200pt]{4.818pt}{0.400pt}}
\put(338.0,429.0){\rule[-0.200pt]{4.818pt}{0.400pt}}
\put(382.0,355.0){\rule[-0.200pt]{0.400pt}{19.031pt}}
\put(372.0,355.0){\rule[-0.200pt]{4.818pt}{0.400pt}}
\put(372.0,434.0){\rule[-0.200pt]{4.818pt}{0.400pt}}
\put(417.0,353.0){\rule[-0.200pt]{0.400pt}{21.922pt}}
\put(407.0,353.0){\rule[-0.200pt]{4.818pt}{0.400pt}}
\put(407.0,444.0){\rule[-0.200pt]{4.818pt}{0.400pt}}
\put(451.0,352.0){\rule[-0.200pt]{0.400pt}{23.849pt}}
\put(441.0,352.0){\rule[-0.200pt]{4.818pt}{0.400pt}}
\put(441.0,451.0){\rule[-0.200pt]{4.818pt}{0.400pt}}
\put(486.0,361.0){\rule[-0.200pt]{0.400pt}{26.017pt}}
\put(476.0,361.0){\rule[-0.200pt]{4.818pt}{0.400pt}}
\put(476.0,469.0){\rule[-0.200pt]{4.818pt}{0.400pt}}
\put(520.0,346.0){\rule[-0.200pt]{0.400pt}{28.185pt}}
\put(510.0,346.0){\rule[-0.200pt]{4.818pt}{0.400pt}}
\put(510.0,463.0){\rule[-0.200pt]{4.818pt}{0.400pt}}
\put(555.0,313.0){\rule[-0.200pt]{0.400pt}{30.835pt}}
\put(545.0,313.0){\rule[-0.200pt]{4.818pt}{0.400pt}}
\put(545.0,441.0){\rule[-0.200pt]{4.818pt}{0.400pt}}
\put(589.0,267.0){\rule[-0.200pt]{0.400pt}{34.449pt}}
\put(579.0,267.0){\rule[-0.200pt]{4.818pt}{0.400pt}}
\put(579.0,410.0){\rule[-0.200pt]{4.818pt}{0.400pt}}
\put(623.0,245.0){\rule[-0.200pt]{0.400pt}{41.676pt}}
\put(613.0,245.0){\rule[-0.200pt]{4.818pt}{0.400pt}}
\put(613.0,418.0){\rule[-0.200pt]{4.818pt}{0.400pt}}
\put(658.0,209.0){\rule[-0.200pt]{0.400pt}{47.939pt}}
\put(648.0,209.0){\rule[-0.200pt]{4.818pt}{0.400pt}}
\put(648.0,408.0){\rule[-0.200pt]{4.818pt}{0.400pt}}
\put(692.0,177.0){\rule[-0.200pt]{0.400pt}{50.348pt}}
\put(682.0,177.0){\rule[-0.200pt]{4.818pt}{0.400pt}}
\put(682.0,386.0){\rule[-0.200pt]{4.818pt}{0.400pt}}
\put(727.0,149.0){\rule[-0.200pt]{0.400pt}{54.202pt}}
\put(717.0,149.0){\rule[-0.200pt]{4.818pt}{0.400pt}}
\put(717.0,374.0){\rule[-0.200pt]{4.818pt}{0.400pt}}
\put(520,419){\usebox{\plotpoint}}
\multiput(520.00,417.92)(0.562,-0.499){303}{\rule{0.550pt}{0.120pt}}
\multiput(520.00,418.17)(170.859,-153.000){2}{\rule{0.275pt}{0.400pt}}
\end{picture}
\hspace*{-0.8in}
\setlength{\unitlength}{0.240900pt}
\ifx\plotpoint\undefined\newsavebox{\plotpoint}\fi
\sbox{\plotpoint}{\rule[-0.200pt]{0.400pt}{0.400pt}}%
\begin{picture}(825,629)(0,0)
\font\gnuplot=cmr10 at 10pt
\gnuplot
\sbox{\plotpoint}{\rule[-0.200pt]{0.400pt}{0.400pt}}%
\put(176.0,168.0){\rule[-0.200pt]{4.818pt}{0.400pt}}
\put(741.0,168.0){\rule[-0.200pt]{4.818pt}{0.400pt}}
\put(176.0,277.0){\rule[-0.200pt]{4.818pt}{0.400pt}}
\put(741.0,277.0){\rule[-0.200pt]{4.818pt}{0.400pt}}
\put(176.0,386.0){\rule[-0.200pt]{4.818pt}{0.400pt}}
\put(741.0,386.0){\rule[-0.200pt]{4.818pt}{0.400pt}}
\put(176.0,495.0){\rule[-0.200pt]{4.818pt}{0.400pt}}
\put(741.0,495.0){\rule[-0.200pt]{4.818pt}{0.400pt}}
\put(176.0,113.0){\rule[-0.200pt]{0.400pt}{4.818pt}}
\put(176,68){\makebox(0,0){0}}
\put(176.0,541.0){\rule[-0.200pt]{0.400pt}{4.818pt}}
\put(348.0,113.0){\rule[-0.200pt]{0.400pt}{4.818pt}}
\put(348,68){\makebox(0,0){5}}
\put(348.0,541.0){\rule[-0.200pt]{0.400pt}{4.818pt}}
\put(520.0,113.0){\rule[-0.200pt]{0.400pt}{4.818pt}}
\put(520,68){\makebox(0,0){10}}
\put(520.0,541.0){\rule[-0.200pt]{0.400pt}{4.818pt}}
\put(692.0,113.0){\rule[-0.200pt]{0.400pt}{4.818pt}}
\put(692,68){\makebox(0,0){15}}
\put(692.0,541.0){\rule[-0.200pt]{0.400pt}{4.818pt}}
\put(176.0,113.0){\rule[-0.200pt]{140.926pt}{0.400pt}}
\put(761.0,113.0){\rule[-0.200pt]{0.400pt}{107.923pt}}
\put(176.0,561.0){\rule[-0.200pt]{140.926pt}{0.400pt}}
\put(468,23){\makebox(0,0){$t_f$}}
\put(468,606){\makebox(0,0){$\kappa_v=0.152$}}
\put(245,277){\makebox(0,0)[l]{{\tiny $M=-0.07(4)$}}}
\put(245,233){\makebox(0,0)[l]{{\tiny $\chi^2=0.05$}}}
\put(176.0,113.0){\rule[-0.200pt]{0.400pt}{107.923pt}}
\put(210,386){\circle*{18}}
\put(245,384){\circle*{18}}
\put(279,384){\circle*{18}}
\put(314,404){\circle*{18}}
\put(348,441){\circle*{18}}
\put(382,430){\circle*{18}}
\put(417,414){\circle*{18}}
\put(451,402){\circle*{18}}
\put(486,396){\circle*{18}}
\put(520,371){\circle*{18}}
\put(555,358){\circle*{18}}
\put(589,346){\circle*{18}}
\put(623,352){\circle*{18}}
\put(658,316){\circle*{18}}
\put(692,261){\circle*{18}}
\put(727,250){\circle*{18}}
\put(210,386){\usebox{\plotpoint}}
\put(200.0,386.0){\rule[-0.200pt]{4.818pt}{0.400pt}}
\put(200.0,386.0){\rule[-0.200pt]{4.818pt}{0.400pt}}
\put(245.0,379.0){\rule[-0.200pt]{0.400pt}{2.409pt}}
\put(235.0,379.0){\rule[-0.200pt]{4.818pt}{0.400pt}}
\put(235.0,389.0){\rule[-0.200pt]{4.818pt}{0.400pt}}
\put(279.0,374.0){\rule[-0.200pt]{0.400pt}{4.818pt}}
\put(269.0,374.0){\rule[-0.200pt]{4.818pt}{0.400pt}}
\put(269.0,394.0){\rule[-0.200pt]{4.818pt}{0.400pt}}
\put(314.0,381.0){\rule[-0.200pt]{0.400pt}{11.081pt}}
\put(304.0,381.0){\rule[-0.200pt]{4.818pt}{0.400pt}}
\put(304.0,427.0){\rule[-0.200pt]{4.818pt}{0.400pt}}
\put(348.0,405.0){\rule[-0.200pt]{0.400pt}{17.586pt}}
\put(338.0,405.0){\rule[-0.200pt]{4.818pt}{0.400pt}}
\put(338.0,478.0){\rule[-0.200pt]{4.818pt}{0.400pt}}
\put(382.0,381.0){\rule[-0.200pt]{0.400pt}{23.608pt}}
\put(372.0,381.0){\rule[-0.200pt]{4.818pt}{0.400pt}}
\put(372.0,479.0){\rule[-0.200pt]{4.818pt}{0.400pt}}
\put(417.0,357.0){\rule[-0.200pt]{0.400pt}{27.703pt}}
\put(407.0,357.0){\rule[-0.200pt]{4.818pt}{0.400pt}}
\put(407.0,472.0){\rule[-0.200pt]{4.818pt}{0.400pt}}
\put(451.0,338.0){\rule[-0.200pt]{0.400pt}{30.835pt}}
\put(441.0,338.0){\rule[-0.200pt]{4.818pt}{0.400pt}}
\put(441.0,466.0){\rule[-0.200pt]{4.818pt}{0.400pt}}
\put(486.0,325.0){\rule[-0.200pt]{0.400pt}{34.449pt}}
\put(476.0,325.0){\rule[-0.200pt]{4.818pt}{0.400pt}}
\put(476.0,468.0){\rule[-0.200pt]{4.818pt}{0.400pt}}
\put(520.0,296.0){\rule[-0.200pt]{0.400pt}{36.135pt}}
\put(510.0,296.0){\rule[-0.200pt]{4.818pt}{0.400pt}}
\put(510.0,446.0){\rule[-0.200pt]{4.818pt}{0.400pt}}
\put(555.0,281.0){\rule[-0.200pt]{0.400pt}{37.339pt}}
\put(545.0,281.0){\rule[-0.200pt]{4.818pt}{0.400pt}}
\put(545.0,436.0){\rule[-0.200pt]{4.818pt}{0.400pt}}
\put(589.0,263.0){\rule[-0.200pt]{0.400pt}{40.230pt}}
\put(579.0,263.0){\rule[-0.200pt]{4.818pt}{0.400pt}}
\put(579.0,430.0){\rule[-0.200pt]{4.818pt}{0.400pt}}
\put(623.0,257.0){\rule[-0.200pt]{0.400pt}{46.012pt}}
\put(613.0,257.0){\rule[-0.200pt]{4.818pt}{0.400pt}}
\put(613.0,448.0){\rule[-0.200pt]{4.818pt}{0.400pt}}
\put(658.0,210.0){\rule[-0.200pt]{0.400pt}{51.312pt}}
\put(648.0,210.0){\rule[-0.200pt]{4.818pt}{0.400pt}}
\put(648.0,423.0){\rule[-0.200pt]{4.818pt}{0.400pt}}
\put(692.0,148.0){\rule[-0.200pt]{0.400pt}{54.443pt}}
\put(682.0,148.0){\rule[-0.200pt]{4.818pt}{0.400pt}}
\put(682.0,374.0){\rule[-0.200pt]{4.818pt}{0.400pt}}
\put(727.0,124.0){\rule[-0.200pt]{0.400pt}{60.466pt}}
\put(717.0,124.0){\rule[-0.200pt]{4.818pt}{0.400pt}}
\put(717.0,375.0){\rule[-0.200pt]{4.818pt}{0.400pt}}
\put(520,384){\usebox{\plotpoint}}
\multiput(520.00,382.92)(0.996,-0.499){205}{\rule{0.896pt}{0.120pt}}
\multiput(520.00,383.17)(205.140,-104.000){2}{\rule{0.448pt}{0.400pt}}
\end{picture}
\hspace*{-0.8in}
\setlength{\unitlength}{0.240900pt}
\ifx\plotpoint\undefined\newsavebox{\plotpoint}\fi
\sbox{\plotpoint}{\rule[-0.200pt]{0.400pt}{0.400pt}}%
\begin{picture}(825,629)(0,0)
\font\gnuplot=cmr10 at 10pt
\gnuplot
\sbox{\plotpoint}{\rule[-0.200pt]{0.400pt}{0.400pt}}%
\put(176.0,168.0){\rule[-0.200pt]{4.818pt}{0.400pt}}
\put(741.0,168.0){\rule[-0.200pt]{4.818pt}{0.400pt}}
\put(176.0,277.0){\rule[-0.200pt]{4.818pt}{0.400pt}}
\put(741.0,277.0){\rule[-0.200pt]{4.818pt}{0.400pt}}
\put(176.0,386.0){\rule[-0.200pt]{4.818pt}{0.400pt}}
\put(741.0,386.0){\rule[-0.200pt]{4.818pt}{0.400pt}}
\put(176.0,495.0){\rule[-0.200pt]{4.818pt}{0.400pt}}
\put(741.0,495.0){\rule[-0.200pt]{4.818pt}{0.400pt}}
\put(176.0,113.0){\rule[-0.200pt]{0.400pt}{4.818pt}}
\put(176,68){\makebox(0,0){0}}
\put(176.0,541.0){\rule[-0.200pt]{0.400pt}{4.818pt}}
\put(348.0,113.0){\rule[-0.200pt]{0.400pt}{4.818pt}}
\put(348,68){\makebox(0,0){5}}
\put(348.0,541.0){\rule[-0.200pt]{0.400pt}{4.818pt}}
\put(520.0,113.0){\rule[-0.200pt]{0.400pt}{4.818pt}}
\put(520,68){\makebox(0,0){10}}
\put(520.0,541.0){\rule[-0.200pt]{0.400pt}{4.818pt}}
\put(692.0,113.0){\rule[-0.200pt]{0.400pt}{4.818pt}}
\put(692,68){\makebox(0,0){15}}
\put(692.0,541.0){\rule[-0.200pt]{0.400pt}{4.818pt}}
\put(176.0,113.0){\rule[-0.200pt]{140.926pt}{0.400pt}}
\put(761.0,113.0){\rule[-0.200pt]{0.400pt}{107.923pt}}
\put(176.0,561.0){\rule[-0.200pt]{140.926pt}{0.400pt}}
\put(468,23){\makebox(0,0){$t_f$}}
\put(468,606){\makebox(0,0){$\kappa_v=0.148$}}
\put(245,277){\makebox(0,0)[l]{{\tiny $M=-0.04(2)$}}}
\put(245,233){\makebox(0,0)[l]{{\tiny $\chi^2=0.1$}}}
\put(176.0,113.0){\rule[-0.200pt]{0.400pt}{107.923pt}}
\put(210,386){\circle*{18}}
\put(245,383){\circle*{18}}
\put(279,383){\circle*{18}}
\put(314,390){\circle*{18}}
\put(348,399){\circle*{18}}
\put(382,394){\circle*{18}}
\put(417,381){\circle*{18}}
\put(451,375){\circle*{18}}
\put(486,365){\circle*{18}}
\put(520,355){\circle*{18}}
\put(555,355){\circle*{18}}
\put(589,358){\circle*{18}}
\put(623,371){\circle*{18}}
\put(658,356){\circle*{18}}
\put(692,354){\circle*{18}}
\put(727,363){\circle*{18}}
\put(210,386){\usebox{\plotpoint}}
\put(200.0,386.0){\rule[-0.200pt]{4.818pt}{0.400pt}}
\put(200.0,386.0){\rule[-0.200pt]{4.818pt}{0.400pt}}
\put(245.0,381.0){\rule[-0.200pt]{0.400pt}{1.204pt}}
\put(235.0,381.0){\rule[-0.200pt]{4.818pt}{0.400pt}}
\put(235.0,386.0){\rule[-0.200pt]{4.818pt}{0.400pt}}
\put(279.0,379.0){\rule[-0.200pt]{0.400pt}{1.927pt}}
\put(269.0,379.0){\rule[-0.200pt]{4.818pt}{0.400pt}}
\put(269.0,387.0){\rule[-0.200pt]{4.818pt}{0.400pt}}
\put(314.0,381.0){\rule[-0.200pt]{0.400pt}{4.336pt}}
\put(304.0,381.0){\rule[-0.200pt]{4.818pt}{0.400pt}}
\put(304.0,399.0){\rule[-0.200pt]{4.818pt}{0.400pt}}
\put(348.0,384.0){\rule[-0.200pt]{0.400pt}{6.986pt}}
\put(338.0,384.0){\rule[-0.200pt]{4.818pt}{0.400pt}}
\put(338.0,413.0){\rule[-0.200pt]{4.818pt}{0.400pt}}
\put(382.0,375.0){\rule[-0.200pt]{0.400pt}{9.154pt}}
\put(372.0,375.0){\rule[-0.200pt]{4.818pt}{0.400pt}}
\put(372.0,413.0){\rule[-0.200pt]{4.818pt}{0.400pt}}
\put(417.0,359.0){\rule[-0.200pt]{0.400pt}{10.840pt}}
\put(407.0,359.0){\rule[-0.200pt]{4.818pt}{0.400pt}}
\put(407.0,404.0){\rule[-0.200pt]{4.818pt}{0.400pt}}
\put(451.0,350.0){\rule[-0.200pt]{0.400pt}{12.045pt}}
\put(441.0,350.0){\rule[-0.200pt]{4.818pt}{0.400pt}}
\put(441.0,400.0){\rule[-0.200pt]{4.818pt}{0.400pt}}
\put(486.0,338.0){\rule[-0.200pt]{0.400pt}{13.009pt}}
\put(476.0,338.0){\rule[-0.200pt]{4.818pt}{0.400pt}}
\put(476.0,392.0){\rule[-0.200pt]{4.818pt}{0.400pt}}
\put(520.0,327.0){\rule[-0.200pt]{0.400pt}{13.731pt}}
\put(510.0,327.0){\rule[-0.200pt]{4.818pt}{0.400pt}}
\put(510.0,384.0){\rule[-0.200pt]{4.818pt}{0.400pt}}
\put(555.0,325.0){\rule[-0.200pt]{0.400pt}{14.454pt}}
\put(545.0,325.0){\rule[-0.200pt]{4.818pt}{0.400pt}}
\put(545.0,385.0){\rule[-0.200pt]{4.818pt}{0.400pt}}
\put(589.0,327.0){\rule[-0.200pt]{0.400pt}{15.177pt}}
\put(579.0,327.0){\rule[-0.200pt]{4.818pt}{0.400pt}}
\put(579.0,390.0){\rule[-0.200pt]{4.818pt}{0.400pt}}
\put(623.0,336.0){\rule[-0.200pt]{0.400pt}{16.863pt}}
\put(613.0,336.0){\rule[-0.200pt]{4.818pt}{0.400pt}}
\put(613.0,406.0){\rule[-0.200pt]{4.818pt}{0.400pt}}
\put(658.0,318.0){\rule[-0.200pt]{0.400pt}{18.308pt}}
\put(648.0,318.0){\rule[-0.200pt]{4.818pt}{0.400pt}}
\put(648.0,394.0){\rule[-0.200pt]{4.818pt}{0.400pt}}
\put(692.0,314.0){\rule[-0.200pt]{0.400pt}{19.272pt}}
\put(682.0,314.0){\rule[-0.200pt]{4.818pt}{0.400pt}}
\put(682.0,394.0){\rule[-0.200pt]{4.818pt}{0.400pt}}
\put(727.0,322.0){\rule[-0.200pt]{0.400pt}{19.513pt}}
\put(717.0,322.0){\rule[-0.200pt]{4.818pt}{0.400pt}}
\put(717.0,403.0){\rule[-0.200pt]{4.818pt}{0.400pt}}
\put(417,381){\usebox{\plotpoint}}
\multiput(417.00,379.92)(2.683,-0.497){49}{\rule{2.223pt}{0.120pt}}
\multiput(417.00,380.17)(133.386,-26.000){2}{\rule{1.112pt}{0.400pt}}
\end{picture}
\]
\caption{The ratio of Eq. (\ref{mff}) as a function of $t_f$ so that the
slope is $G_{M,{\rm dis}}(q^2)$ at $\vec{q} = 2\pi/La$. The sea quark is fixed
at $\kappa_s = 0.154$, the strange quark mass, and the valence quark
masses are at $\kappa_v = 0.148$, 0.152, and 0.154. M is the fitted slope.}
\end{figure}
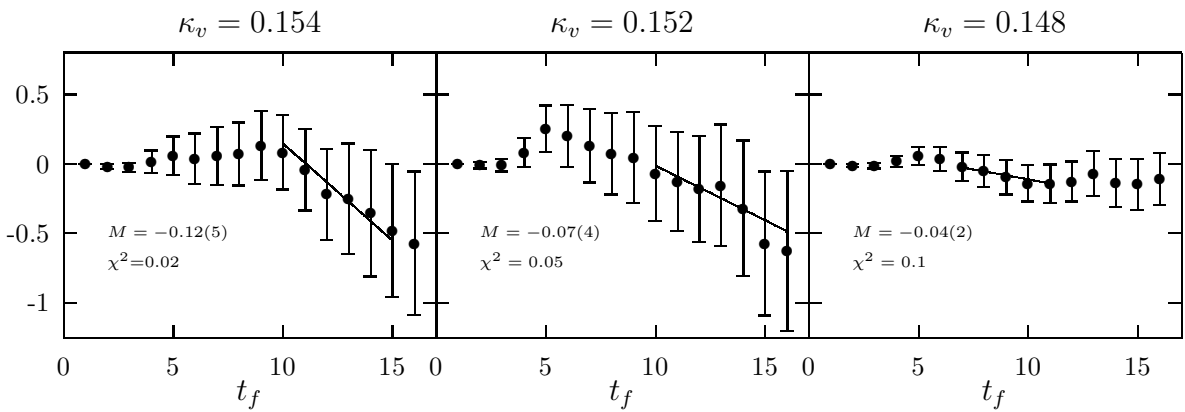

Plotted in Fig. 2(a) is the extrapolated $G_M^s(q^2)$ at 4 nonzero $q^2$
values.
The errors are again obtained by jackknifing the extrapolation procedure
with the covariance matrix used to include the correlation among the
three valence $\kappa$'s. In view of the fact that the scalar current
exhibits a very soft form factor for the sea quark
(i.e. $g_{S,{\rm dis}} (q^2)$)
which has been fitted well with a monopole form~\cite{dll96}, we shall
similarly use a monopole form to extrapolate $G_M^s(q^2)$ with nonzero
$q^2$ to $G_M^s(0)$. Indicated as $\circ$ in Fig. 2(a), we find
$G_M^s(0) = - 0.36 \pm 0.20 $. Again, the correlation
among the 4 $q^2$ are taken into account and the error is from
jackknifing the fitting procedure. This is consistent with the
recent experimental value within errors (see Table 1).
The monopole mass is found to be $0.58 \pm 0.16$ of $m_N$.
To explore the uncertainty of the $q^2$ dependence, we also fitted
$G_M^s(q^2)$ with a dipole form and found $G_M^s(0) = - 0.27 \pm 0.12$
with a dipole mass $m_D/m_N = 1.19 \pm 0.22$.
Similar results are obtained for u and d quarks with monopole fits.
They turn out to be $G_{M,{\rm dis}}^{u/d}(0) =
 -0.65 \pm 0.30$, which is about 1.8 times the size of $G_M^s(0)$.
These are tabulated in Table 1.
 
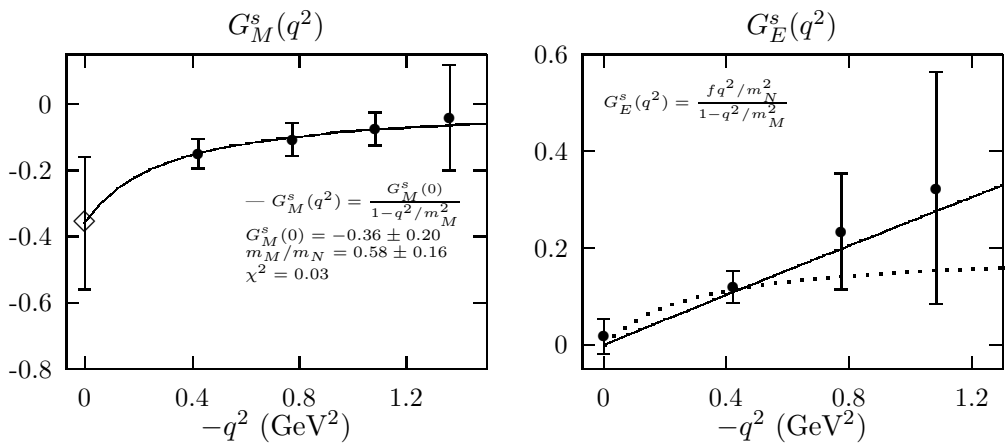
\begin{figure}[h]
\[
\hspace*{-1.2in}
\setlength{\unitlength}{0.240900pt}
\ifx\plotpoint\undefined\newsavebox{\plotpoint}\fi
\sbox{\plotpoint}{\rule[-0.200pt]{0.400pt}{0.400pt}}%
\begin{picture}(900,675)(0,0)
\font\gnuplot=cmr10 at 10pt
\gnuplot
\sbox{\plotpoint}{\rule[-0.200pt]{0.400pt}{0.400pt}}%
\put(176.0,113.0){\rule[-0.200pt]{4.818pt}{0.400pt}}
\put(154,113){\makebox(0,0)[r]{-0.8}}
\put(816.0,113.0){\rule[-0.200pt]{4.818pt}{0.400pt}}
\put(176.0,217.0){\rule[-0.200pt]{4.818pt}{0.400pt}}
\put(154,217){\makebox(0,0)[r]{-0.6}}
\put(816.0,217.0){\rule[-0.200pt]{4.818pt}{0.400pt}}
\put(176.0,321.0){\rule[-0.200pt]{4.818pt}{0.400pt}}
\put(154,321){\makebox(0,0)[r]{-0.4}}
\put(816.0,321.0){\rule[-0.200pt]{4.818pt}{0.400pt}}
\put(176.0,425.0){\rule[-0.200pt]{4.818pt}{0.400pt}}
\put(154,425){\makebox(0,0)[r]{-0.2}}
\put(816.0,425.0){\rule[-0.200pt]{4.818pt}{0.400pt}}
\put(176.0,529.0){\rule[-0.200pt]{4.818pt}{0.400pt}}
\put(154,529){\makebox(0,0)[r]{0}}
\put(816.0,529.0){\rule[-0.200pt]{4.818pt}{0.400pt}}
\put(205.0,113.0){\rule[-0.200pt]{0.400pt}{4.818pt}}
\put(205,68){\makebox(0,0){0}}
\put(205.0,587.0){\rule[-0.200pt]{0.400pt}{4.818pt}}
\put(374.0,113.0){\rule[-0.200pt]{0.400pt}{4.818pt}}
\put(374,68){\makebox(0,0){0.4}}
\put(374.0,587.0){\rule[-0.200pt]{0.400pt}{4.818pt}}
\put(542.0,113.0){\rule[-0.200pt]{0.400pt}{4.818pt}}
\put(542,68){\makebox(0,0){0.8}}
\put(542.0,587.0){\rule[-0.200pt]{0.400pt}{4.818pt}}
\put(710.0,113.0){\rule[-0.200pt]{0.400pt}{4.818pt}}
\put(710,68){\makebox(0,0){1.2}}
\put(710.0,587.0){\rule[-0.200pt]{0.400pt}{4.818pt}}
\put(176.0,113.0){\rule[-0.200pt]{158.994pt}{0.400pt}}
\put(836.0,113.0){\rule[-0.200pt]{0.400pt}{119.005pt}}
\put(176.0,607.0){\rule[-0.200pt]{158.994pt}{0.400pt}}
\put(506,23){\makebox(0,0){{\small $-q^2$ (GeV$^2$)}}}
\put(506,652){\makebox(0,0){{\small $G_M^s(q^2)$}}}
\put(458,373){\makebox(0,0)[l]{{\tiny ---  $G_M^s(q^2)=\frac{G_M^s(0)}{1-q^2/m_M^2}$}}}
\put(458,321){\makebox(0,0)[l]{{\tiny $G_M^s(0)=-0.36\pm 0.20$}}}
\put(458,295){\makebox(0,0)[l]{{\tiny $m_M/m_N = 0.58\pm 0.16$}}}
\put(458,264){\makebox(0,0)[l]{{\tiny $\chi^2=0.03$}}}
\put(176.0,113.0){\rule[-0.200pt]{0.400pt}{119.005pt}}
\put(383,451){\circle*{18}}
\put(531,474){\circle*{18}}
\put(661,490){\circle*{18}}
\put(778,508){\circle*{18}}
\put(383.0,428.0){\rule[-0.200pt]{0.400pt}{11.081pt}}
\put(373.0,428.0){\rule[-0.200pt]{4.818pt}{0.400pt}}
\put(373.0,474.0){\rule[-0.200pt]{4.818pt}{0.400pt}}
\put(531.0,448.0){\rule[-0.200pt]{0.400pt}{12.527pt}}
\put(521.0,448.0){\rule[-0.200pt]{4.818pt}{0.400pt}}
\put(521.0,500.0){\rule[-0.200pt]{4.818pt}{0.400pt}}
\put(661.0,464.0){\rule[-0.200pt]{0.400pt}{12.527pt}}
\put(651.0,464.0){\rule[-0.200pt]{4.818pt}{0.400pt}}
\put(651.0,516.0){\rule[-0.200pt]{4.818pt}{0.400pt}}
\put(778.0,425.0){\rule[-0.200pt]{0.400pt}{39.989pt}}
\put(768.0,425.0){\rule[-0.200pt]{4.818pt}{0.400pt}}
\put(768.0,591.0){\rule[-0.200pt]{4.818pt}{0.400pt}}
\put(205,342){\usebox{\plotpoint}}
\multiput(205.58,342.00)(0.497,0.615){49}{\rule{0.120pt}{0.592pt}}
\multiput(204.17,342.00)(26.000,30.771){2}{\rule{0.400pt}{0.296pt}}
\multiput(231.00,374.58)(0.564,0.496){43}{\rule{0.552pt}{0.120pt}}
\multiput(231.00,373.17)(24.854,23.000){2}{\rule{0.276pt}{0.400pt}}
\multiput(257.00,397.58)(0.768,0.495){31}{\rule{0.712pt}{0.119pt}}
\multiput(257.00,396.17)(24.523,17.000){2}{\rule{0.356pt}{0.400pt}}
\multiput(283.00,414.58)(0.972,0.493){23}{\rule{0.869pt}{0.119pt}}
\multiput(283.00,413.17)(23.196,13.000){2}{\rule{0.435pt}{0.400pt}}
\multiput(308.00,427.58)(1.329,0.491){17}{\rule{1.140pt}{0.118pt}}
\multiput(308.00,426.17)(23.634,10.000){2}{\rule{0.570pt}{0.400pt}}
\multiput(334.00,437.59)(1.485,0.489){15}{\rule{1.256pt}{0.118pt}}
\multiput(334.00,436.17)(23.394,9.000){2}{\rule{0.628pt}{0.400pt}}
\multiput(360.00,446.59)(1.942,0.485){11}{\rule{1.586pt}{0.117pt}}
\multiput(360.00,445.17)(22.709,7.000){2}{\rule{0.793pt}{0.400pt}}
\multiput(386.00,453.59)(2.208,0.482){9}{\rule{1.767pt}{0.116pt}}
\multiput(386.00,452.17)(21.333,6.000){2}{\rule{0.883pt}{0.400pt}}
\multiput(411.00,459.59)(2.825,0.477){7}{\rule{2.180pt}{0.115pt}}
\multiput(411.00,458.17)(21.475,5.000){2}{\rule{1.090pt}{0.400pt}}
\multiput(437.00,464.60)(3.698,0.468){5}{\rule{2.700pt}{0.113pt}}
\multiput(437.00,463.17)(20.396,4.000){2}{\rule{1.350pt}{0.400pt}}
\multiput(463.00,468.60)(3.698,0.468){5}{\rule{2.700pt}{0.113pt}}
\multiput(463.00,467.17)(20.396,4.000){2}{\rule{1.350pt}{0.400pt}}
\multiput(489.00,472.61)(5.374,0.447){3}{\rule{3.433pt}{0.108pt}}
\multiput(489.00,471.17)(17.874,3.000){2}{\rule{1.717pt}{0.400pt}}
\multiput(514.00,475.61)(5.597,0.447){3}{\rule{3.567pt}{0.108pt}}
\multiput(514.00,474.17)(18.597,3.000){2}{\rule{1.783pt}{0.400pt}}
\multiput(540.00,478.61)(5.597,0.447){3}{\rule{3.567pt}{0.108pt}}
\multiput(540.00,477.17)(18.597,3.000){2}{\rule{1.783pt}{0.400pt}}
\multiput(566.00,481.61)(5.597,0.447){3}{\rule{3.567pt}{0.108pt}}
\multiput(566.00,480.17)(18.597,3.000){2}{\rule{1.783pt}{0.400pt}}
\put(592,484.17){\rule{5.100pt}{0.400pt}}
\multiput(592.00,483.17)(14.415,2.000){2}{\rule{2.550pt}{0.400pt}}
\put(617,486.17){\rule{5.300pt}{0.400pt}}
\multiput(617.00,485.17)(15.000,2.000){2}{\rule{2.650pt}{0.400pt}}
\put(643,488.17){\rule{5.300pt}{0.400pt}}
\multiput(643.00,487.17)(15.000,2.000){2}{\rule{2.650pt}{0.400pt}}
\put(669,489.67){\rule{6.263pt}{0.400pt}}
\multiput(669.00,489.17)(13.000,1.000){2}{\rule{3.132pt}{0.400pt}}
\put(695,491.17){\rule{5.100pt}{0.400pt}}
\multiput(695.00,490.17)(14.415,2.000){2}{\rule{2.550pt}{0.400pt}}
\put(720,492.67){\rule{6.263pt}{0.400pt}}
\multiput(720.00,492.17)(13.000,1.000){2}{\rule{3.132pt}{0.400pt}}
\put(746,493.67){\rule{6.263pt}{0.400pt}}
\multiput(746.00,493.17)(13.000,1.000){2}{\rule{3.132pt}{0.400pt}}
\put(772,495.17){\rule{5.300pt}{0.400pt}}
\multiput(772.00,494.17)(15.000,2.000){2}{\rule{2.650pt}{0.400pt}}
\put(798,496.67){\rule{6.023pt}{0.400pt}}
\multiput(798.00,496.17)(12.500,1.000){2}{\rule{3.011pt}{0.400pt}}
\put(823.0,498.0){\rule[-0.200pt]{3.132pt}{0.400pt}}
\put(205,342){\raisebox{-.8pt}{\makebox(0,0){$\Diamond$}}}
\put(205.0,238.0){\rule[-0.200pt]{0.400pt}{50.107pt}}
\put(195.0,238.0){\rule[-0.200pt]{4.818pt}{0.400pt}}
\put(195.0,446.0){\rule[-0.200pt]{4.818pt}{0.400pt}}
\end{picture}
\hspace*{-0.3in}
\setlength{\unitlength}{0.240900pt}
\ifx\plotpoint\undefined\newsavebox{\plotpoint}\fi
\sbox{\plotpoint}{\rule[-0.200pt]{0.400pt}{0.400pt}}%
\begin{picture}(900,675)(0,0)
\font\gnuplot=cmr10 at 10pt
\gnuplot
\sbox{\plotpoint}{\rule[-0.200pt]{0.400pt}{0.400pt}}%
\put(176.0,151.0){\rule[-0.200pt]{4.818pt}{0.400pt}}
\put(154,151){\makebox(0,0)[r]{0}}
\put(816.0,151.0){\rule[-0.200pt]{4.818pt}{0.400pt}}
\put(176.0,303.0){\rule[-0.200pt]{4.818pt}{0.400pt}}
\put(154,303){\makebox(0,0)[r]{0.2}}
\put(816.0,303.0){\rule[-0.200pt]{4.818pt}{0.400pt}}
\put(176.0,455.0){\rule[-0.200pt]{4.818pt}{0.400pt}}
\put(154,455){\makebox(0,0)[r]{0.4}}
\put(816.0,455.0){\rule[-0.200pt]{4.818pt}{0.400pt}}
\put(176.0,607.0){\rule[-0.200pt]{4.818pt}{0.400pt}}
\put(154,607){\makebox(0,0)[r]{0.6}}
\put(816.0,607.0){\rule[-0.200pt]{4.818pt}{0.400pt}}
\put(210.0,113.0){\rule[-0.200pt]{0.400pt}{4.818pt}}
\put(210,68){\makebox(0,0){0}}
\put(210.0,587.0){\rule[-0.200pt]{0.400pt}{4.818pt}}
\put(402.0,113.0){\rule[-0.200pt]{0.400pt}{4.818pt}}
\put(402,68){\makebox(0,0){0.4}}
\put(402.0,587.0){\rule[-0.200pt]{0.400pt}{4.818pt}}
\put(595.0,113.0){\rule[-0.200pt]{0.400pt}{4.818pt}}
\put(595,68){\makebox(0,0){0.8}}
\put(595.0,587.0){\rule[-0.200pt]{0.400pt}{4.818pt}}
\put(788.0,113.0){\rule[-0.200pt]{0.400pt}{4.818pt}}
\put(788,68){\makebox(0,0){1.2}}
\put(788.0,587.0){\rule[-0.200pt]{0.400pt}{4.818pt}}
\put(176.0,113.0){\rule[-0.200pt]{158.994pt}{0.400pt}}
\put(836.0,113.0){\rule[-0.200pt]{0.400pt}{119.005pt}}
\put(176.0,607.0){\rule[-0.200pt]{158.994pt}{0.400pt}}
\put(506,23){\makebox(0,0){{\small $-q^2$ (GeV$^2$)}}}
\put(506,652){\makebox(0,0){{\small $G_E^s(q^2)$}}}
\put(210,531){\makebox(0,0)[l]{{\tiny $G_E^s(q^2)=\frac{f q^2/m_N^2}{1-q^2/m_M^2}$}}}
\put(176.0,113.0){\rule[-0.200pt]{0.400pt}{119.005pt}}
\put(210,165){\circle*{18}}
\put(413,242){\circle*{18}}
\put(583,329){\circle*{18}}
\put(732,397){\circle*{18}}
\put(210.0,137.0){\rule[-0.200pt]{0.400pt}{13.249pt}}
\put(200.0,137.0){\rule[-0.200pt]{4.818pt}{0.400pt}}
\put(200.0,192.0){\rule[-0.200pt]{4.818pt}{0.400pt}}
\put(413.0,217.0){\rule[-0.200pt]{0.400pt}{12.045pt}}
\put(403.0,217.0){\rule[-0.200pt]{4.818pt}{0.400pt}}
\put(403.0,267.0){\rule[-0.200pt]{4.818pt}{0.400pt}}
\put(583.0,238.0){\rule[-0.200pt]{0.400pt}{43.844pt}}
\put(573.0,238.0){\rule[-0.200pt]{4.818pt}{0.400pt}}
\put(573.0,420.0){\rule[-0.200pt]{4.818pt}{0.400pt}}
\put(732.0,215.0){\rule[-0.200pt]{0.400pt}{87.928pt}}
\put(722.0,215.0){\rule[-0.200pt]{4.818pt}{0.400pt}}
\put(722.0,580.0){\rule[-0.200pt]{4.818pt}{0.400pt}}
\sbox{\plotpoint}{\rule[-0.500pt]{1.000pt}{1.000pt}}%
\put(210,151){\usebox{\plotpoint}}
\put(210.00,151.00){\usebox{\plotpoint}}
\put(224.77,165.57){\usebox{\plotpoint}}
\put(241.53,177.81){\usebox{\plotpoint}}
\put(258.88,189.13){\usebox{\plotpoint}}
\put(277.32,198.66){\usebox{\plotpoint}}
\multiput(282,201)(19.077,8.176){0}{\usebox{\plotpoint}}
\put(296.27,207.11){\usebox{\plotpoint}}
\put(315.70,214.34){\usebox{\plotpoint}}
\put(335.52,220.51){\usebox{\plotpoint}}
\multiput(340,222)(19.957,5.702){0}{\usebox{\plotpoint}}
\put(355.44,226.29){\usebox{\plotpoint}}
\put(375.77,230.45){\usebox{\plotpoint}}
\put(396.10,234.62){\usebox{\plotpoint}}
\multiput(398,235)(20.295,4.349){0}{\usebox{\plotpoint}}
\put(416.42,238.88){\usebox{\plotpoint}}
\put(436.86,242.41){\usebox{\plotpoint}}
\multiput(441,243)(20.547,2.935){0}{\usebox{\plotpoint}}
\put(457.41,245.32){\usebox{\plotpoint}}
\put(477.98,248.14){\usebox{\plotpoint}}
\put(498.64,249.98){\usebox{\plotpoint}}
\multiput(499,250)(20.547,2.935){0}{\usebox{\plotpoint}}
\put(519.24,252.42){\usebox{\plotpoint}}
\put(539.85,254.69){\usebox{\plotpoint}}
\multiput(542,255)(20.710,1.381){0}{\usebox{\plotpoint}}
\put(560.54,256.25){\usebox{\plotpoint}}
\put(581.17,258.45){\usebox{\plotpoint}}
\multiput(585,259)(20.710,1.381){0}{\usebox{\plotpoint}}
\put(601.85,260.13){\usebox{\plotpoint}}
\put(622.55,261.57){\usebox{\plotpoint}}
\multiput(629,262)(20.703,1.479){0}{\usebox{\plotpoint}}
\put(643.26,263.02){\usebox{\plotpoint}}
\put(663.98,264.00){\usebox{\plotpoint}}
\put(684.71,264.85){\usebox{\plotpoint}}
\multiput(687,265)(20.703,1.479){0}{\usebox{\plotpoint}}
\put(705.41,266.29){\usebox{\plotpoint}}
\put(726.12,267.72){\usebox{\plotpoint}}
\multiput(730,268)(20.756,0.000){0}{\usebox{\plotpoint}}
\put(746.86,268.19){\usebox{\plotpoint}}
\put(767.59,269.00){\usebox{\plotpoint}}
\multiput(773,269)(20.710,1.381){0}{\usebox{\plotpoint}}
\put(788.31,270.02){\usebox{\plotpoint}}
\put(809.03,271.00){\usebox{\plotpoint}}
\put(829.75,271.91){\usebox{\plotpoint}}
\multiput(831,272)(20.756,0.000){0}{\usebox{\plotpoint}}
\put(836,272){\usebox{\plotpoint}}
\sbox{\plotpoint}{\rule[-0.200pt]{0.400pt}{0.400pt}}%
\put(210,151){\usebox{\plotpoint}}
\multiput(210.00,151.59)(1.220,0.488){13}{\rule{1.050pt}{0.117pt}}
\multiput(210.00,150.17)(16.821,8.000){2}{\rule{0.525pt}{0.400pt}}
\multiput(229.00,159.59)(1.220,0.488){13}{\rule{1.050pt}{0.117pt}}
\multiput(229.00,158.17)(16.821,8.000){2}{\rule{0.525pt}{0.400pt}}
\multiput(248.00,167.59)(1.286,0.488){13}{\rule{1.100pt}{0.117pt}}
\multiput(248.00,166.17)(17.717,8.000){2}{\rule{0.550pt}{0.400pt}}
\multiput(268.00,175.59)(1.220,0.488){13}{\rule{1.050pt}{0.117pt}}
\multiput(268.00,174.17)(16.821,8.000){2}{\rule{0.525pt}{0.400pt}}
\multiput(287.00,183.59)(1.220,0.488){13}{\rule{1.050pt}{0.117pt}}
\multiput(287.00,182.17)(16.821,8.000){2}{\rule{0.525pt}{0.400pt}}
\multiput(306.00,191.59)(1.408,0.485){11}{\rule{1.186pt}{0.117pt}}
\multiput(306.00,190.17)(16.539,7.000){2}{\rule{0.593pt}{0.400pt}}
\multiput(325.00,198.59)(1.286,0.488){13}{\rule{1.100pt}{0.117pt}}
\multiput(325.00,197.17)(17.717,8.000){2}{\rule{0.550pt}{0.400pt}}
\multiput(345.00,206.59)(1.220,0.488){13}{\rule{1.050pt}{0.117pt}}
\multiput(345.00,205.17)(16.821,8.000){2}{\rule{0.525pt}{0.400pt}}
\multiput(364.00,214.59)(1.220,0.488){13}{\rule{1.050pt}{0.117pt}}
\multiput(364.00,213.17)(16.821,8.000){2}{\rule{0.525pt}{0.400pt}}
\multiput(383.00,222.59)(1.220,0.488){13}{\rule{1.050pt}{0.117pt}}
\multiput(383.00,221.17)(16.821,8.000){2}{\rule{0.525pt}{0.400pt}}
\multiput(402.00,230.59)(1.286,0.488){13}{\rule{1.100pt}{0.117pt}}
\multiput(402.00,229.17)(17.717,8.000){2}{\rule{0.550pt}{0.400pt}}
\multiput(422.00,238.59)(1.408,0.485){11}{\rule{1.186pt}{0.117pt}}
\multiput(422.00,237.17)(16.539,7.000){2}{\rule{0.593pt}{0.400pt}}
\multiput(441.00,245.59)(1.220,0.488){13}{\rule{1.050pt}{0.117pt}}
\multiput(441.00,244.17)(16.821,8.000){2}{\rule{0.525pt}{0.400pt}}
\multiput(460.00,253.59)(1.286,0.488){13}{\rule{1.100pt}{0.117pt}}
\multiput(460.00,252.17)(17.717,8.000){2}{\rule{0.550pt}{0.400pt}}
\multiput(480.00,261.59)(1.220,0.488){13}{\rule{1.050pt}{0.117pt}}
\multiput(480.00,260.17)(16.821,8.000){2}{\rule{0.525pt}{0.400pt}}
\multiput(499.00,269.59)(1.408,0.485){11}{\rule{1.186pt}{0.117pt}}
\multiput(499.00,268.17)(16.539,7.000){2}{\rule{0.593pt}{0.400pt}}
\multiput(518.00,276.59)(1.220,0.488){13}{\rule{1.050pt}{0.117pt}}
\multiput(518.00,275.17)(16.821,8.000){2}{\rule{0.525pt}{0.400pt}}
\multiput(537.00,284.59)(1.286,0.488){13}{\rule{1.100pt}{0.117pt}}
\multiput(537.00,283.17)(17.717,8.000){2}{\rule{0.550pt}{0.400pt}}
\multiput(557.00,292.59)(1.408,0.485){11}{\rule{1.186pt}{0.117pt}}
\multiput(557.00,291.17)(16.539,7.000){2}{\rule{0.593pt}{0.400pt}}
\multiput(576.00,299.59)(1.220,0.488){13}{\rule{1.050pt}{0.117pt}}
\multiput(576.00,298.17)(16.821,8.000){2}{\rule{0.525pt}{0.400pt}}
\multiput(595.00,307.59)(1.220,0.488){13}{\rule{1.050pt}{0.117pt}}
\multiput(595.00,306.17)(16.821,8.000){2}{\rule{0.525pt}{0.400pt}}
\multiput(614.00,315.59)(1.484,0.485){11}{\rule{1.243pt}{0.117pt}}
\multiput(614.00,314.17)(17.420,7.000){2}{\rule{0.621pt}{0.400pt}}
\multiput(634.00,322.59)(1.220,0.488){13}{\rule{1.050pt}{0.117pt}}
\multiput(634.00,321.17)(16.821,8.000){2}{\rule{0.525pt}{0.400pt}}
\multiput(653.00,330.59)(1.220,0.488){13}{\rule{1.050pt}{0.117pt}}
\multiput(653.00,329.17)(16.821,8.000){2}{\rule{0.525pt}{0.400pt}}
\multiput(672.00,338.59)(1.408,0.485){11}{\rule{1.186pt}{0.117pt}}
\multiput(672.00,337.17)(16.539,7.000){2}{\rule{0.593pt}{0.400pt}}
\multiput(691.00,345.59)(1.286,0.488){13}{\rule{1.100pt}{0.117pt}}
\multiput(691.00,344.17)(17.717,8.000){2}{\rule{0.550pt}{0.400pt}}
\multiput(711.00,353.59)(1.220,0.488){13}{\rule{1.050pt}{0.117pt}}
\multiput(711.00,352.17)(16.821,8.000){2}{\rule{0.525pt}{0.400pt}}
\multiput(730.00,361.59)(1.408,0.485){11}{\rule{1.186pt}{0.117pt}}
\multiput(730.00,360.17)(16.539,7.000){2}{\rule{0.593pt}{0.400pt}}
\multiput(749.00,368.59)(1.286,0.488){13}{\rule{1.100pt}{0.117pt}}
\multiput(749.00,367.17)(17.717,8.000){2}{\rule{0.550pt}{0.400pt}}
\multiput(769.00,376.59)(1.408,0.485){11}{\rule{1.186pt}{0.117pt}}
\multiput(769.00,375.17)(16.539,7.000){2}{\rule{0.593pt}{0.400pt}}
\multiput(788.00,383.59)(1.220,0.488){13}{\rule{1.050pt}{0.117pt}}
\multiput(788.00,382.17)(16.821,8.000){2}{\rule{0.525pt}{0.400pt}}
\multiput(807.00,391.59)(1.408,0.485){11}{\rule{1.186pt}{0.117pt}}
\multiput(807.00,390.17)(16.539,7.000){2}{\rule{0.593pt}{0.400pt}}
\multiput(826.00,398.60)(1.358,0.468){5}{\rule{1.100pt}{0.113pt}}
\multiput(826.00,397.17)(7.717,4.000){2}{\rule{0.550pt}{0.400pt}}
\end{picture}
\]
\caption{(a) Strange magnetic form factor $G_M^s(q^2)$. $G_M^s(0)$, indicated
by $\diamond$, is obtained from a monopole fit. (b) Strangeness electric
form factor $G_E^s(q^2)$. The solid line is a fit with the monopole form
shown in the figure and the dashed line is obtained with the monopole mass
$m_M$ from $G_M^s(q^2)$ in (a).}
\end{figure}
 
Now, we are ready to address the question of why the SU(6) relation is badly
broken in the scalar current (e.g. $F_S$, $D_S$) and axial current
(e.g. $g_A^0$) and yet is so good for the neutron-proton magnetic moment
ratio $\mu_n/\mu_p$. The lattice calculations for the scalar~\cite{dll96}
and axial~\cite{dll95} currents reveal the fact that the SU(6) breaking
comes from both the sea quarks in the DI and the cloud
quarks in the CI. We shall see how these degrees of
freedom play out in the case of the m.\,m. We first plot in Fig. 3 the ratio
$\mu_n/\mu_p$ for the CI part (shown as $\circ$) as a function of the valence
quark mass.
We see that when the quark mass is near the charm region ($m_q a$ at 0.55
corresponds to $m_q \sim 1$ GeV), the ratio is close to the SU(6) prediction
of -2/3.
This is quite reasonable as this is in the non-relativistic regime where one
expects SU(6) to work well. As the quark mass comes down to the strange
region ($m_qa =0.07$), the ratio becomes less negative. Extrapolated to the
chiral limit, the ratio is $ - 0.616 \pm 0.022$ which deviates from the SU(6)
prediction by 8 \%. We understand this deviation as mainly due to the
cloud quark effect in the
Z-graphs. As we switch off these Z-graphs in a valence approximation
~\cite{dll95,dll96}, the ratio (plotted as $\diamond$ in Fig. 3)
becomes closer to the SU(6) value which resembles the non-relativistic
case. Similar behaviors were observed for the scalar and axial matrix
elements~\cite{dll96,dll95}. Now we add the sea quark contribution from the
DI to give $\mu_{dis} = (2/3 G_{M,{\rm dis}}^u(0) -1/3
G_{M,{\rm dis}}^d(0) -1/3 G_M^s(0))
\mu_N $ to the CI and find that it tends to cancel the cloud effect and
bring the ratio back to be similar to what
the valence approximation predicts. For the
$G_M^s(0)$ at various $\kappa_v$, we use the $G_M^s(0)/G_{M,{\rm dis}}^u(0)$
ratio
from the chiral limit to obtain it from the $G_{M,{\rm dis}}^u(0)$ at each
$\kappa_v$. At the chiral limit, when the total sea contribution
$\mu_{dis} = -0.097 \pm 0.037 \mu_N$ is added to the CI,
the $\mu_n/\mu_p$ ratio then comes down to $-0.68 \pm 0.04$
which is consistent with the experimental value of 0.685.
We note that the $\mu_n/\mu_p$ ratio for the
full result ($\bullet$) is more negative at the chiral limit compared
with those at other $m_qa$.  This has to do with the fact that the CI employs
the linear quark mass extrapolation, as do other observables for the
CI~\cite{dll95,dll96,ldd95}, whereas the DI uses the $\sqrt{m_q}$ dependence
for the chiral extrapolation as mentioned above. From this analysis,
we see that although the individual $G_{M,{\rm dis}}^u(0),
G_{M,{\rm dis}}^d(0)$, and $G_M^s(0)$ are large, their net contribution
$\mu_{\rm dis} = -0.097 \pm 0.037\mu_N$ to $\mu_n$ and $\mu_p$ is much smaller
because of the partial cancellation due to the quark charges of u, d, and s.
The sea contribution turns out to be further canceled by the cloud effect
to bring the $\mu_n/\mu_p$ ratio close to the experimental value and
the SU(6) relation. Barring any known symmetry principle yet to
surface, this cancellation is probably accidental and in stark contrast
with the $\pi N \sigma$ term and flavor-singlet $g_A^0$ where the
cloud and sea effects add up to enhance the SU(6) breaking~\cite{dll96,
dll95}.
 
\begin{figure}[h]
\setlength{\unitlength}{0.240900pt}
\ifx\plotpoint\undefined\newsavebox{\plotpoint}\fi
\sbox{\plotpoint}{\rule[-0.200pt]{0.400pt}{0.400pt}}%
\begin{picture}(1800,900)(0,0)
\font\gnuplot=cmr10 at 10pt
\gnuplot
\sbox{\plotpoint}{\rule[-0.200pt]{0.400pt}{0.400pt}}%
\put(176.0,173.0){\rule[-0.200pt]{4.818pt}{0.400pt}}
\put(154,173){\makebox(0,0)[r]{-0.75}}
\put(1716.0,173.0){\rule[-0.200pt]{4.818pt}{0.400pt}}
\put(176.0,293.0){\rule[-0.200pt]{4.818pt}{0.400pt}}
\put(154,293){\makebox(0,0)[r]{-0.7}}
\put(1716.0,293.0){\rule[-0.200pt]{4.818pt}{0.400pt}}
\put(176.0,413.0){\rule[-0.200pt]{4.818pt}{0.400pt}}
\put(154,413){\makebox(0,0)[r]{-0.65}}
\put(1716.0,413.0){\rule[-0.200pt]{4.818pt}{0.400pt}}
\put(176.0,532.0){\rule[-0.200pt]{4.818pt}{0.400pt}}
\put(154,532){\makebox(0,0)[r]{-0.6}}
\put(1716.0,532.0){\rule[-0.200pt]{4.818pt}{0.400pt}}
\put(176.0,652.0){\rule[-0.200pt]{4.818pt}{0.400pt}}
\put(154,652){\makebox(0,0)[r]{-0.55}}
\put(1716.0,652.0){\rule[-0.200pt]{4.818pt}{0.400pt}}
\put(176.0,772.0){\rule[-0.200pt]{4.818pt}{0.400pt}}
\put(154,772){\makebox(0,0)[r]{-0.5}}
\put(1716.0,772.0){\rule[-0.200pt]{4.818pt}{0.400pt}}
\put(238.0,113.0){\rule[-0.200pt]{0.400pt}{4.818pt}}
\put(238,68){\makebox(0,0){0}}
\put(238.0,812.0){\rule[-0.200pt]{0.400pt}{4.818pt}}
\put(488.0,113.0){\rule[-0.200pt]{0.400pt}{4.818pt}}
\put(488,68){\makebox(0,0){0.1}}
\put(488.0,812.0){\rule[-0.200pt]{0.400pt}{4.818pt}}
\put(738.0,113.0){\rule[-0.200pt]{0.400pt}{4.818pt}}
\put(738,68){\makebox(0,0){0.2}}
\put(738.0,812.0){\rule[-0.200pt]{0.400pt}{4.818pt}}
\put(987.0,113.0){\rule[-0.200pt]{0.400pt}{4.818pt}}
\put(987,68){\makebox(0,0){0.3}}
\put(987.0,812.0){\rule[-0.200pt]{0.400pt}{4.818pt}}
\put(1237.0,113.0){\rule[-0.200pt]{0.400pt}{4.818pt}}
\put(1237,68){\makebox(0,0){0.4}}
\put(1237.0,812.0){\rule[-0.200pt]{0.400pt}{4.818pt}}
\put(1486.0,113.0){\rule[-0.200pt]{0.400pt}{4.818pt}}
\put(1486,68){\makebox(0,0){0.5}}
\put(1486.0,812.0){\rule[-0.200pt]{0.400pt}{4.818pt}}
\put(1736.0,113.0){\rule[-0.200pt]{0.400pt}{4.818pt}}
\put(1736,68){\makebox(0,0){0.6}}
\put(1736.0,812.0){\rule[-0.200pt]{0.400pt}{4.818pt}}
\put(176.0,113.0){\rule[-0.200pt]{375.804pt}{0.400pt}}
\put(1736.0,113.0){\rule[-0.200pt]{0.400pt}{173.207pt}}
\put(176.0,832.0){\rule[-0.200pt]{375.804pt}{0.400pt}}
\put(956,-22){\makebox(0,0){{\bf $m_q a$}}}
\put(956,877){\makebox(0,0){{\small $\mu_n/\mu_p$}}}
\put(176.0,113.0){\rule[-0.200pt]{0.400pt}{173.207pt}}
\put(1362,700){\makebox(0,0)[r]{{\scriptsize Connected Insertion}}}
\put(1406,700){\circle{24}}
\put(238,494){\circle{24}}
\put(410,489){\circle{24}}
\put(531,470){\circle{24}}
\put(776,441){\circle{24}}
\put(1222,434){\circle{24}}
\put(1590,410){\circle{24}}
\put(1384.0,700.0){\rule[-0.200pt]{15.899pt}{0.400pt}}
\put(1384.0,690.0){\rule[-0.200pt]{0.400pt}{4.818pt}}
\put(1450.0,690.0){\rule[-0.200pt]{0.400pt}{4.818pt}}
\put(238.0,441.0){\rule[-0.200pt]{0.400pt}{25.535pt}}
\put(228.0,441.0){\rule[-0.200pt]{4.818pt}{0.400pt}}
\put(228.0,547.0){\rule[-0.200pt]{4.818pt}{0.400pt}}
\put(410.0,432.0){\rule[-0.200pt]{0.400pt}{27.703pt}}
\put(400.0,432.0){\rule[-0.200pt]{4.818pt}{0.400pt}}
\put(400.0,547.0){\rule[-0.200pt]{4.818pt}{0.400pt}}
\put(531.0,417.0){\rule[-0.200pt]{0.400pt}{25.535pt}}
\put(521.0,417.0){\rule[-0.200pt]{4.818pt}{0.400pt}}
\put(521.0,523.0){\rule[-0.200pt]{4.818pt}{0.400pt}}
\put(776.0,393.0){\rule[-0.200pt]{0.400pt}{23.126pt}}
\put(766.0,393.0){\rule[-0.200pt]{4.818pt}{0.400pt}}
\put(766.0,489.0){\rule[-0.200pt]{4.818pt}{0.400pt}}
\put(1222.0,427.0){\rule[-0.200pt]{0.400pt}{3.373pt}}
\put(1212.0,427.0){\rule[-0.200pt]{4.818pt}{0.400pt}}
\put(1212.0,441.0){\rule[-0.200pt]{4.818pt}{0.400pt}}
\put(1590.0,398.0){\rule[-0.200pt]{0.400pt}{5.782pt}}
\put(1580.0,398.0){\rule[-0.200pt]{4.818pt}{0.400pt}}
\put(1580.0,422.0){\rule[-0.200pt]{4.818pt}{0.400pt}}
\put(1362,655){\makebox(0,0)[r]{{\scriptsize Valence}}}
\put(1406,655){\raisebox{-.8pt}{\makebox(0,0){$\Diamond$}}}
\put(445,384){\raisebox{-.8pt}{\makebox(0,0){$\Diamond$}}}
\put(474,381){\raisebox{-.8pt}{\makebox(0,0){$\Diamond$}}}
\put(502,396){\raisebox{-.8pt}{\makebox(0,0){$\Diamond$}}}
\put(616,403){\raisebox{-.8pt}{\makebox(0,0){$\Diamond$}}}
\put(1384.0,655.0){\rule[-0.200pt]{15.899pt}{0.400pt}}
\put(1384.0,645.0){\rule[-0.200pt]{0.400pt}{4.818pt}}
\put(1450.0,645.0){\rule[-0.200pt]{0.400pt}{4.818pt}}
\put(445.0,338.0){\rule[-0.200pt]{0.400pt}{21.922pt}}
\put(435.0,338.0){\rule[-0.200pt]{4.818pt}{0.400pt}}
\put(435.0,429.0){\rule[-0.200pt]{4.818pt}{0.400pt}}
\put(474.0,338.0){\rule[-0.200pt]{0.400pt}{20.958pt}}
\put(464.0,338.0){\rule[-0.200pt]{4.818pt}{0.400pt}}
\put(464.0,425.0){\rule[-0.200pt]{4.818pt}{0.400pt}}
\put(502.0,360.0){\rule[-0.200pt]{0.400pt}{17.345pt}}
\put(492.0,360.0){\rule[-0.200pt]{4.818pt}{0.400pt}}
\put(492.0,432.0){\rule[-0.200pt]{4.818pt}{0.400pt}}
\put(616.0,367.0){\rule[-0.200pt]{0.400pt}{17.345pt}}
\put(606.0,367.0){\rule[-0.200pt]{4.818pt}{0.400pt}}
\put(606.0,439.0){\rule[-0.200pt]{4.818pt}{0.400pt}}
\sbox{\plotpoint}{\rule[-0.500pt]{1.000pt}{1.000pt}}%
\put(1362,610){\makebox(0,0)[r]{{\scriptsize -2/3}}}
\multiput(1384,610)(20.756,0.000){4}{\usebox{\plotpoint}}
\put(1450,610){\usebox{\plotpoint}}
\put(176,373){\usebox{\plotpoint}}
\multiput(176,373)(20.756,0.000){76}{\usebox{\plotpoint}}
\put(1736,373){\usebox{\plotpoint}}
\sbox{\plotpoint}{\rule[-0.200pt]{0.400pt}{0.400pt}}%
\put(1362,565){\makebox(0,0)[r]{{\scriptsize Expt.~~-0.685}}}
\put(1384.0,565.0){\rule[-0.200pt]{15.899pt}{0.400pt}}
\put(176,329){\usebox{\plotpoint}}
\put(176.0,329.0){\rule[-0.200pt]{375.804pt}{0.400pt}}
\put(1362,520){\makebox(0,0)[r]{{\scriptsize Sea + CI}}}
\put(1406,520){\circle*{24}}
\put(238,322){\circle*{24}}
\put(410,427){\circle*{24}}
\put(531,437){\circle*{24}}
\put(776,437){\circle*{24}}
\put(1384.0,520.0){\rule[-0.200pt]{15.899pt}{0.400pt}}
\put(1384.0,510.0){\rule[-0.200pt]{0.400pt}{4.818pt}}
\put(1450.0,510.0){\rule[-0.200pt]{0.400pt}{4.818pt}}
\put(238.0,233.0){\rule[-0.200pt]{0.400pt}{42.639pt}}
\put(228.0,233.0){\rule[-0.200pt]{4.818pt}{0.400pt}}
\put(228.0,410.0){\rule[-0.200pt]{4.818pt}{0.400pt}}
\put(410.0,379.0){\rule[-0.200pt]{0.400pt}{23.126pt}}
\put(400.0,379.0){\rule[-0.200pt]{4.818pt}{0.400pt}}
\put(400.0,475.0){\rule[-0.200pt]{4.818pt}{0.400pt}}
\put(531.0,401.0){\rule[-0.200pt]{0.400pt}{17.345pt}}
\put(521.0,401.0){\rule[-0.200pt]{4.818pt}{0.400pt}}
\put(521.0,473.0){\rule[-0.200pt]{4.818pt}{0.400pt}}
\put(776.0,413.0){\rule[-0.200pt]{0.400pt}{11.563pt}}
\put(766.0,413.0){\rule[-0.200pt]{4.818pt}{0.400pt}}
\put(766.0,461.0){\rule[-0.200pt]{4.818pt}{0.400pt}}
\end{picture}
\caption{Neutron to proton m.\,m. ratio $\mu_n/\mu_p$ as a function of
the dimensionless quark mass $m_qa$. The $\diamond$
indicates the valence result. The $\circ$ is the result for the CI and
$\bullet$ indicates the full result with the inclusion of the sea from DI.}
\end{figure}
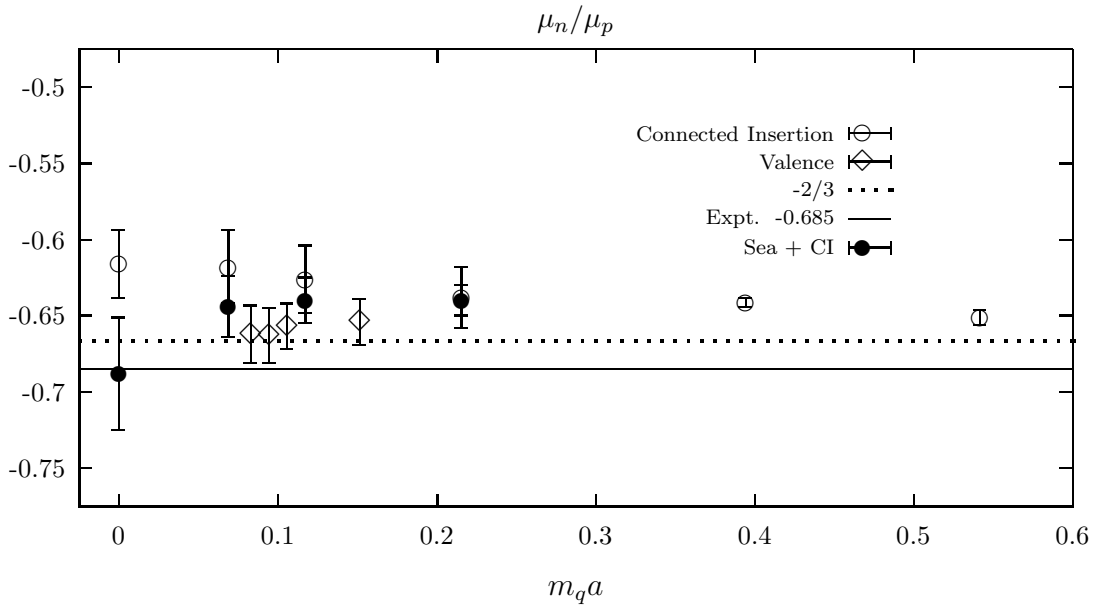

 The sea contribution from the u, d, and s quarks $G_{M,{\rm dis}}^{u,d}(q^2)$
and $G_M^s (q^2)$
are added to the valence and cloud part in the CI, $G_{M,{\rm con}}(q^2)$,
to give the full $G_M^p(q^2)$ and $G_M^n(q^2)$. They are plotted in Fig. 4(a)
and 4(b) and indicated by $\bullet$. Also plotted are the
$G_{M,{\rm con}}(q^2)$
(denoted by $\circ$) and the experimental fits (in
solid line). We see from Fig. 4 and Table 1 that $\mu_p$ and $\mu_n$
are smaller than the experimental results by $\sim$ 6\% in absolute values.
This is presumably due to the systematic errors of the finite
volume, finite lattice spacing, and the quenched approximation.
We should point out that in the earlier discussion of the neuton to
proton m.\,m. ratio $\mu_n/\mu_p$, the systematic errors are expected to be
cancelled out in the ratio to a large extent. Our conclusion of the ratio
$\mu_n/\mu_p$ in the preceding  paragraph is thus based on this assumption.
 
\begin{figure}[h]
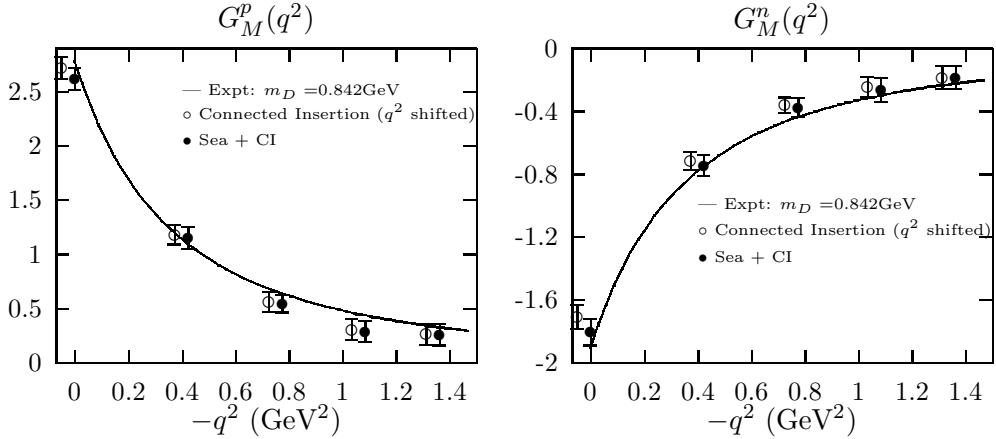

\[
\hspace*{-1.2in}\input{fig4_1.tex}
\hspace*{-0.3in}\input{fig4_2.tex}
\]
\caption{(a) Proton magnetic form factor $G_M^p(q^2)$. $\circ$ indicates
the result from the CI. They are shifted slighted to the left in $ - q^2$
to avoid overlap with the full result which is shown as
$\bullet$. The solid line is the fit to the experiment [24].
(b) the same as in (a) for the neutron form factor $G_M^n(q^2)$.}
\end{figure}

   A similar analysis is done for the strange Sachs electric form factor
$G_E^s(q^2)$. This is plotted in Fig. 2 (b). We see that $G_E^s(0)$ is
consistent with zero as it should be. This serves as a test of the stochastic
noise estimation with the $Z_2$ noise. We fitted $G_E^s(q^2)$ with the form
$G_E^s(q^2) = f\frac{q^2}{m_N^2}/(1 - q^2/m_M^2)$ (solid line in Fig. 2(b)).
 The resultant electric mean-square radius
$\langle r_s^2 \rangle_E = 6 \frac{dG_E^s(q^2)}{dq^2}|_{q^2 = 0} =
- 0.061 \pm 0.003\,{\rm fm}^2$. This is shown in Table 1.
 
In view of the large errors, we also plot the above form for $G_E^s(q^2)$
with the monopole mass $m_M$ taken from $G_M^s(q^2)$ which is shown by the
dashed line. This gives $\langle r_s^2 \rangle_E = - 0.16 \pm 0.06\,{\rm fm}^2$
with $\chi^2/N_{DF}= 0.24$.  This shows that
that the uncertainty in the fitting can be as large
as a factor of two. Nevertheless, $\langle r_s^2 \rangle_E$ is relatively
small. This small negative value in $\langle r_s^2 \rangle_E$ and
large negative $G_M^s(0)$ are consistent with the kaon loop picture
~\cite{mb94} and VMD~\cite{fnj94} but is inconsistent with most of the
other model predictions~\cite{lei96,cbk96}.

   Since the DI of $u$ and $d$ quarks are slighter larger than that
of the $s$ quark, the total sea contribution
$G_{E,{\rm dis}}(q^2) = 2/3 G_{E,{\rm dis}}^u(q^2)
-1/3 G_{E,{\rm dis}}^d(q^2) -1/3 G_E^s(q^2)$ adds a small positive value to the
valence and cloud part $G_{E,{\rm con}}(q^2)$ in the CI.
The proton $G_E^p(q^2)$ and neutron $G_E^n(q^2)$
are plotted in Figs. 5(a) and 5(b) respectively. We see that the CI part
$G_{E,{\rm con}}^p(q^2)$ (shown as $\circ$ in Fig. 5(a)) gives the main
contribution in proton.  $G_{E,{\rm dis}}(q^2)$ adds only a little change to
it. The resultant dipole fit
gives a dipole mass of $0.857 \pm 0.031$ GeV (Table 1). This is consistent
with the experimental dipole mass of 0.842 GeV. In the case of
the neutron,
since $G_{E,{\rm con}}^n(q^2)$~\cite{wdl92} itself ($\circ$ in Fig. 5(b))
is small,
the sea contribution $G_{E,{\rm dis}}(q^2)$ becomes a sizable part of the total
$G_E^n(q^2)$ ($\bullet$ in Fig. 5(b)). We see that when the sea is included
we have a reasonably good match with the experimental results (solid line
in Fig. 5(b)). The total mean square charge radius of $ - 0.123 \pm 0.019
{\rm fm}^2$
is obtained from fitting with the form $G_E^n(q^2) =
f\frac{q^2}{4 m_N^2}/(1 - q^2/M_D^2)^2/(1- 5.6 q^2 /4 M_N^2)$ which has
been used to fit the experimental results~\cite{gal71}. This is consistent
with the experimentally fitted result of $ - 0.127 {\rm fm}^2$.
 
\begin{figure}[h]
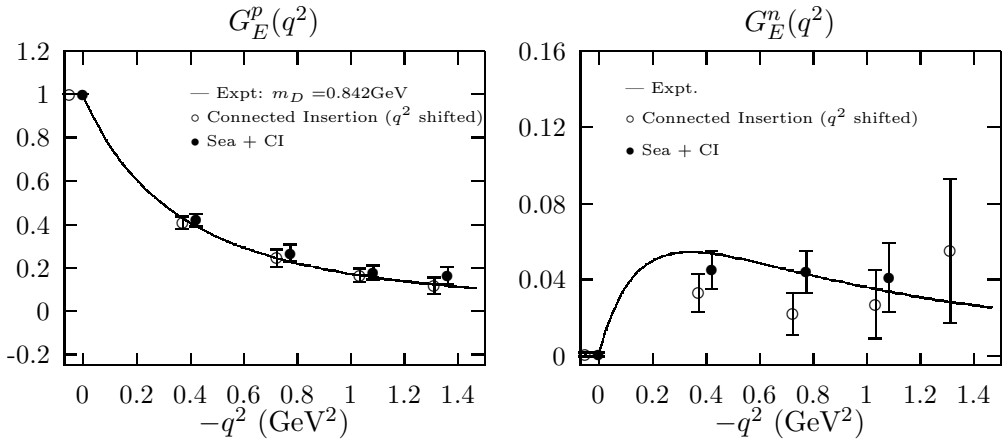

\[
\hspace*{-1.2in}\input{fig5_1.tex}
\hspace*{-0.3in}\input{fig5_2.tex}
\]
\caption{(a) Proton electric form factor $G_E^p(q^2)$. $\circ$ shows
the result from the CI. They are shifted slighted to the left in $ - q^2$
to avoid overlap with the full result which is shown as
$\bullet$. The solid line is the fit to the experiment [24]. (b)
the same as (a) for the neutron form factor $G_E^n(q^2)$.}
\end{figure}

 \begin{table}[ht]
\caption{Strangeness and proton-neutron m.\,m. and charge radii in
comparison with experiments.}
\begin{tabular}{llc}
 \multicolumn{1}{c}{} &\multicolumn{1}{c}{Lattice}
 & \multicolumn{1}{c} {Experiments} \\
 \hline
 $G_M^s(0)$ & $- 0.36 \pm 0.20 $ &
 $G^s_M(Q^2=0.1$GeV$^2)= 0.23\pm 0.37\pm 0.15\pm 0.19$~\cite{SAMPLE97} \\
 $G_{M,{\rm dis}}^u(0) $  & $- 0.65 \pm 0.30 $  &  \\
 $\mu_{{\rm dis}}$ & $- 0.097 \pm 0.037 \mu_N$  &   \\
 $\mu_p$ & $2.62 \pm 0.07\,\mu_N$  & $2.79\, \mu_N$ \\
 $\mu_n$  & $- 1.81 \pm 0.07\, \mu_N$ & $- 1.91\, \mu_N$  \\
 $\mu_n/\mu_p$ & $- 0.68 \pm 0.04$ & $- 0.685$ \\
 $\langle r_s^2 \rangle_E$ &  $- 0.061(3) --- - 0.16(6)\, {\rm fm}^2$ &  \\
 $\langle r^2 \rangle_E^p$ &  $0.636\pm 0.046\,{\rm fm}^2$
 & 0.659 ${\rm fm}^2$~\cite{gal71}\\
 $\langle r^2 \rangle_E^n$ &  $- 0.123 \pm 0.019\,{\rm fm}^2$
 &  $- 0.127 \,{\rm fm}^2$~\cite{gal71}
 \\
 \hline
 \end{tabular}
 \end{table}

In summary, we have calculated the $s$ and $u$, $d$ contributions to
the electric and magnetic form factors of the nucleon. The individual m.\,m.
and electric form factors from the different flavors in the sea are not
small, however there are
large cancellations among themselves due to the electric charges of the
$u$, $d$, and $s$ quarks. We find
that a negative $G_M^s(0)$ leads to a total negative sea contribution to
the nucleon m.\,m. which cancels the cloud effect to make the
$\mu_n/\mu_p$ ratio consistent with the experiment.
We also find $G_E^s(q^2)$ positive and leads to a postive
total sea contribution to the neutron electric form factor $G_E^n(q^2)$.
Future calculations are needed to investigate the systematic errors
associated with the finite volume and lattice spacing as well as the
quenched approximation.

\flushleft{\bf Acknowledgements} \\
 
This work is partially supported by DOE
Grant DE-FG05-84ER40154 and by the Australian Research Council.
The authors wish to thank W. Korsch, D. Leinweber, R. McKeown, and W. Wilcox
for helpful comments.

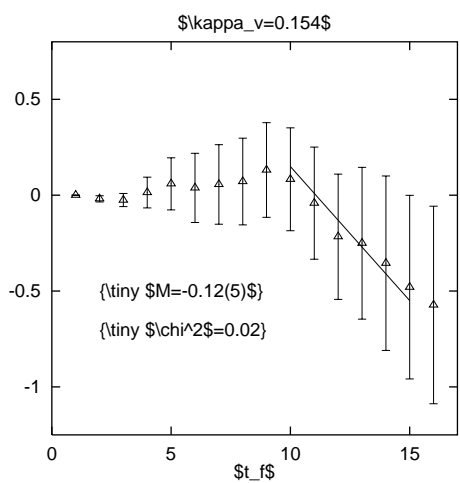
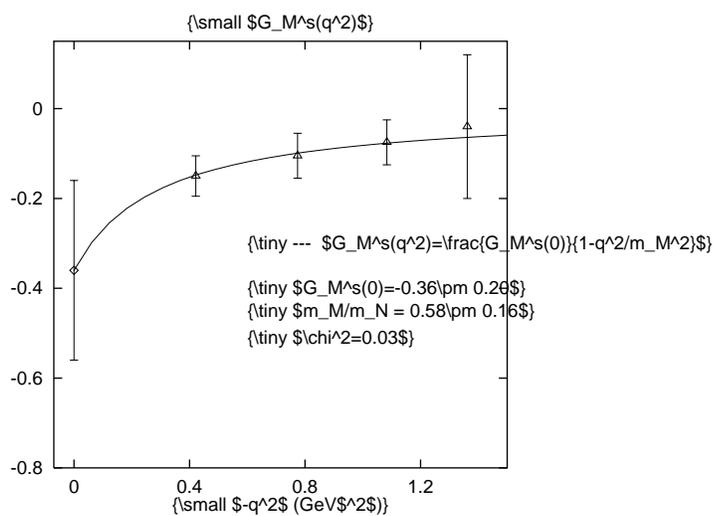
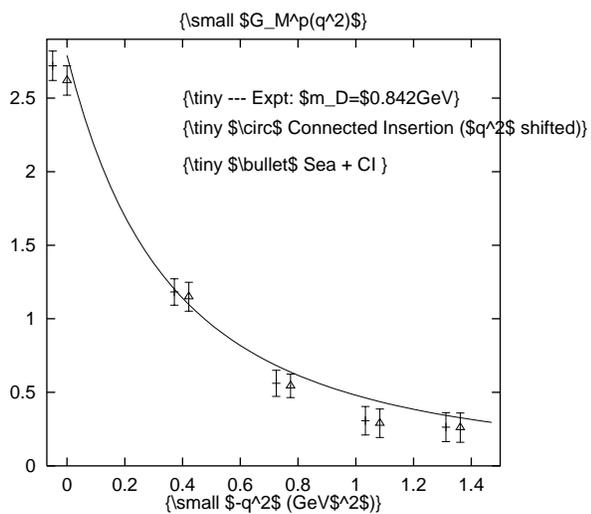
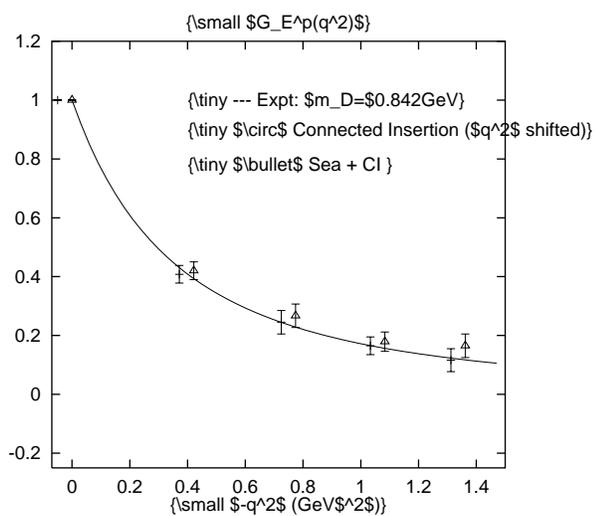
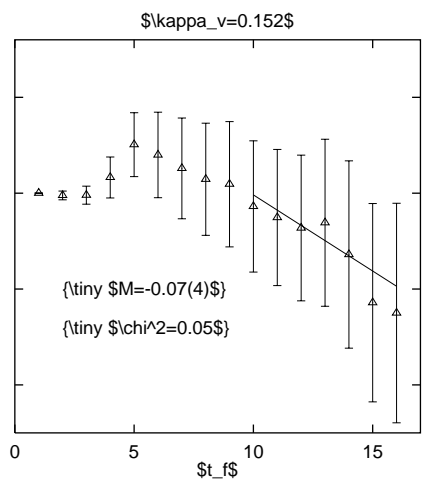
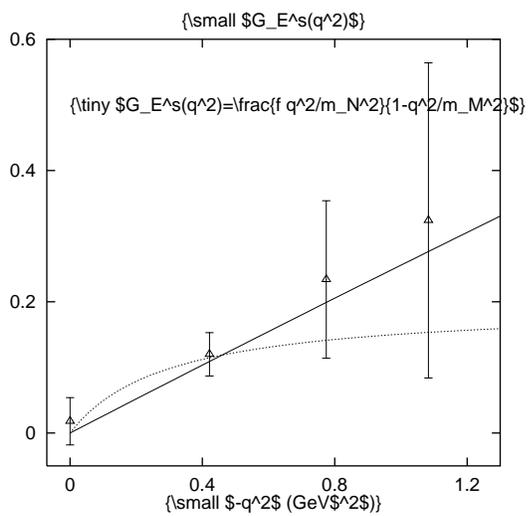
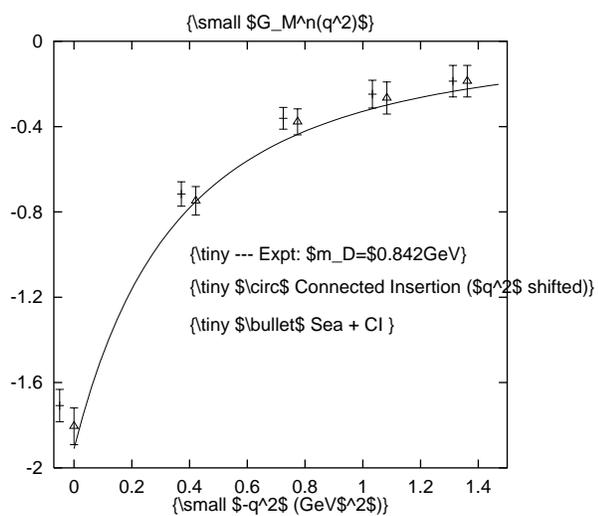
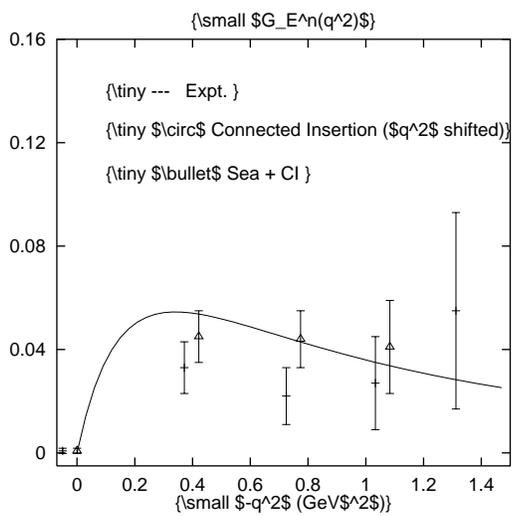
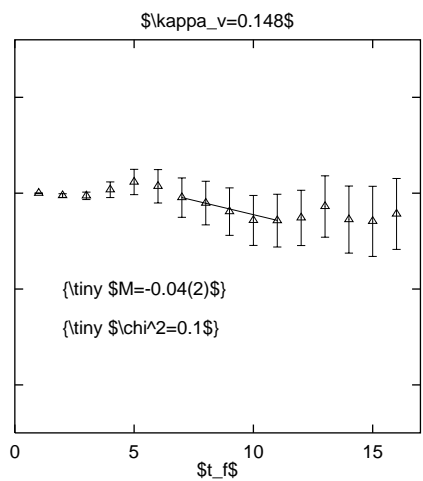
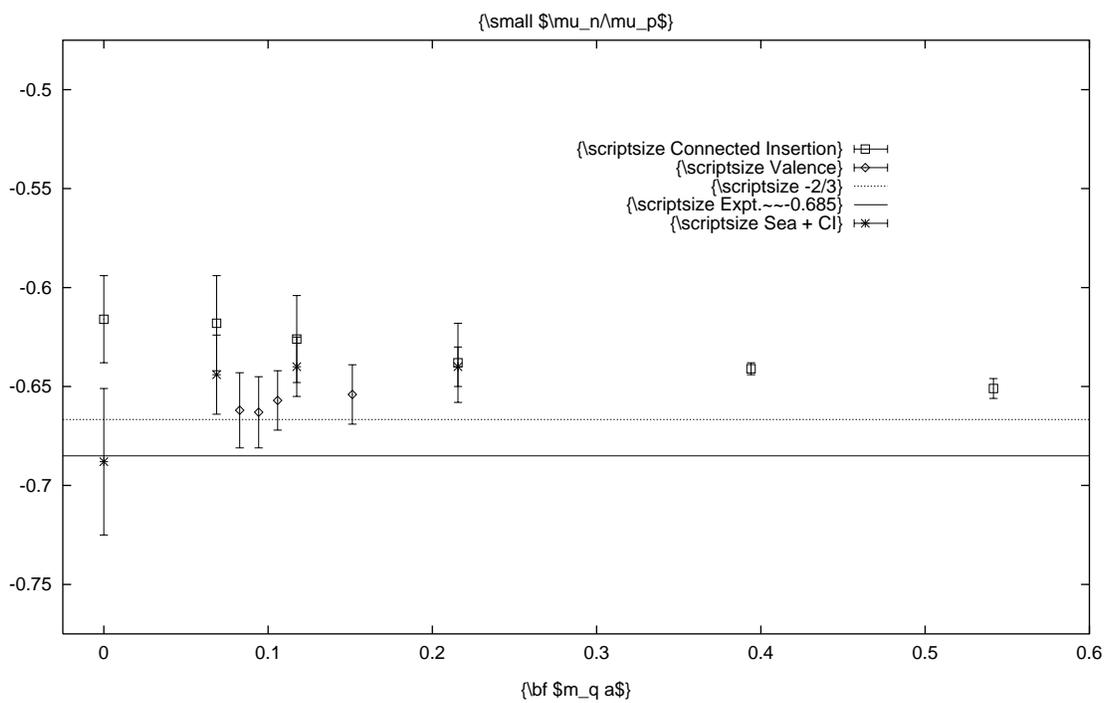
\end{document}